\documentclass[]{aa}
\usepackage{graphicx}
\graphicspath{{./figures/}}
\usepackage{natbib}
\usepackage{amssymb,amsmath,amsfonts}
\usepackage{epsfig}
\usepackage{mathptmx}
\usepackage{longtable}
\usepackage{comment}
\usepackage{color}
\usepackage[caption=false]{subfig}
\usepackage{animate}
\usepackage{rotating}
\usepackage{pdflscape}
\usepackage{url}
\usepackage{enumitem}
\usepackage[draft]{hyperref}
\usepackage[normalem]{ulem}
\def\kMS/{{k\sc{ms}}}
\def\DDF/{{{\sc ddf}acet}}
\def\PSF/{{\sc psf}}
\def\SSD/{{\sc ssd}}
\def\SSDGA/{{\sc ssd-ga}}
\def\LOTTSpipe/{{{LoTSS}-DR1}}
\def\CLEAN/{{\sc clean}}

\begin{document} 

\title{The LOFAR Two-metre Sky Survey -- First Data Release}
\title{The LOFAR Two-metre Sky Survey\thanks{LoTSS}} 
\subtitle{II. First data release}

\authorrunning{Shimwell et~al.}
\titlerunning{The LOFAR Two-metre Sky Survey -- DR1}
\author{T. W. Shimwell$^{1,2}$\thanks{E-mail: shimwell@astron.nl}, 
C. Tasse$^{3,4}$,
M. J. Hardcastle$^{5}$, 
A. P. Mechev$^{2}$, 
W. L. Williams$^{5}$,
P. N. Best$^{6}$,
H. J. A. R\"{o}ttgering$^{2}$, 
J. R. Callingham$^{1}$,
T. J. Dijkema$^{1}$, 
F. de Gasperin$^{2,7}$
% ...
\\
(Affiliations can be found after the references)}
\institute{}
\date{Accepted 12 September 2018; received 04 June 2018; in original form \today}

\abstract{
\noindent
The LOFAR Two-metre Sky Survey (LoTSS) is an ongoing sensitive,
high-resolution 120-168\,MHz survey of the entire northern sky for which observations are now
20\% complete. We present our first full-quality public data
release. For this data release 424 square degrees, or 2\% of the eventual coverage, in the region of the HETDEX Spring Field
(right ascension 10h45m00s to 15h30m00s and declination
45$^\circ$00$\arcmin$00$\arcsec$ to 57$^\circ$00$\arcmin$00$\arcsec$)
were mapped using a fully automated direction-dependent
calibration and imaging pipeline that we developed. A total of
325,694 sources are detected with a signal of at least five times the
noise, and the source density is a factor of $\sim 10$ higher than the
most sensitive existing very wide-area radio-continuum surveys. The
median sensitivity is S$_{\rm 144 MHz} = 71\,\mu$Jy beam$^{-1}$ and the point-source completeness is 90\% at an integrated flux density of 0.45\,mJy. The resolution of the images is 6\,$\arcsec$ and the positional accuracy is within 0.2\,$\arcsec$. This data release consists of a catalogue containing location, flux, and shape estimates together with 58 mosaic images that cover the catalogued area. In this paper we provide an overview of the data release with a focus on the processing of the LOFAR data and the characteristics of the resulting images. In two accompanying papers we provide the radio source associations and deblending and, where possible, the optical identifications of the radio sources together with the photometric redshifts and properties of the host galaxies. These data release papers are published together with a further $\sim$20 articles that highlight the scientific potential of LoTSS.
}

\keywords{surveys -- catalogues -- radio continuum: general -- techniques: image processing}
 \maketitle

\section{Introduction}

\begin{figure}   \centering
   \includegraphics[width=\linewidth]{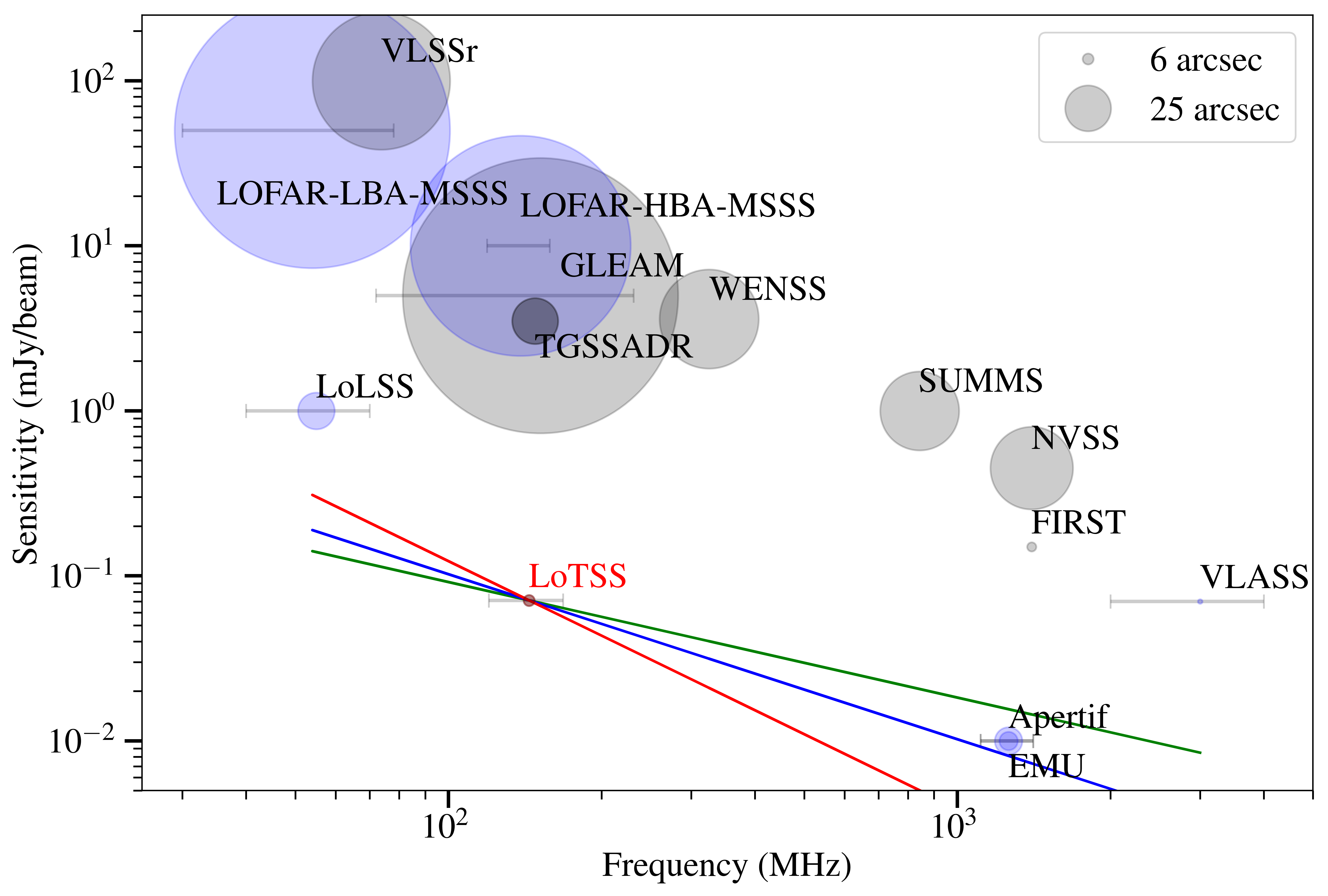}
   \caption{Image rms, frequency, and angular resolution (linearly proportional to the radius of the markers) of LoTSS-DR1 in comparison to a selection of existing wide-area completed (grey) and upcoming (blue) radio surveys. The horizontal lines show the frequency coverage for surveys with large fractional bandwidths. The green, blue, and red lines show an equivalent sensitivity to LoTSS for compact radio sources with spectral indices of -0.7, -1.0, and -1.5, respectively.} 
   \label{fig:lotss_summary}
\end{figure}

Surveys that probe deeply into new parameter space have enormous
discovery potential. The LOFAR Two-metre Sky Survey (LoTSS;
\citealt{Shimwell_2017}) is one example: it is an ongoing survey that is exploiting the unique capabilities of the LOw Frequency ARray (LOFAR; \citealt{vanHaarlem_2013}) to produce a sensitive,
high-resolution radio survey of the northern sky with a frequency coverage
of 120-168\,MHz (see Fig.\ \ref{fig:lotss_summary}). The survey was primarily motivated by the potential of low-frequency observations to facilitate breakthroughs in research areas such as the formation and evolution of massive black  holes (e.g. \citealt{Wilman_2008} and \citealt{Best_2014}) and clusters of galaxies (e.g. \citealt{Cassano_2010} and \citealt{Brunetti_2014}). However, there are many other important scientific drivers of the survey, and there is active research in areas such as high redshift radio sources (e.g. \citealt{Saxena_2017}), galaxy clusters (e.g. \citealt{Botteon_2018}, \citealt{Hoang_2017}, \citealt{degasperin_2017}, \citealt{Savini_2018} and \citealt{Wilber_2018}), active galactic nuclei (e.g. \citealt{Brienza_2017}, \citealt{Morabito_2017} and \citealt{Williams_2018a}), star forming galaxies (e.g. \citealt{Calistro_2017}), gravitational lensing, galactic radio emission, cosmological studies (\citealt{Raccanelli_2012}), magnetic fields (e.g. \citealt{vanEck_2018}), transients and recombination lines (e.g. \citealt{Oonk_2017}).

The LoTSS survey is one of several ongoing or recently completed very wide-area low-frequency
radio surveys that are providing important scientific and
technical insights. Other such surveys include the Multifrequency Snapshot Sky Survey
(MSSS; \citealt{Heald_2015}), TIFR GMRT Sky Survey alternative data
release (TGSS-ADR1; \citealt{Intema_2017}), GaLactic and Extragalactic
All-sky MWA (GLEAM; \citealt{Wayth_2015} and \citealt{Hurley-walker_2017}), LOFAR Low-band Sky Survey
(LoLSS; \citealt{degasperin_2017b}), and the Very Large Array
Low-frequency Sky Survey Redux (VLSSr; \citealt{Lane_2014}). However, LoTSS is designed to push further into new territory. This survey aims to
provide a low-frequency survey that will remain competitive even once the 
Square Kilometre Array (\citealt{Dewdney_2009}) is fully operational,
and will not
be surpassed as a low-frequency wide-area northern sky survey for the foreseeable future. The LoTSS can provide the
astrometric precision that is required for robust identification of optical counterparts (see e.g. \citealt{McAlpine_2012}) and a sensitivity that, for typical radio sources, exceeds that
achieved in existing very wide area higher
frequency surveys such as the NRAO VLA Sky Survey (NVSS; \citealt{Condon_1998}), Faint Images of the Radio Sky at
Twenty-Centimeters (FIRST; \citealt{Becker_1995}), Sydney University Molonglo Sky Survey (SUMSS; \citealt{Bock_1999,Mauch_2003}),  and WEsterbork
Northern Sky Survey (WENSS; \citealt{Rengelink_1997}) and rivals
forthcoming higher frequency surveys such as the Evolutionary
Map of the Universe (EMU; \citealt{Norris_2011}), the APERture Tile In Focus survey (e.g. \citealt{Rottgering_2011})
and the VLA Sky Survey
(VLASS\footnote{\url{https://science.nrao.edu/science/surveys/vlass}}). More
specifically the primary observational objectives of LoTSS are to
reach a sensitivity of less than 100\,$\mu$Jy beam$^{-1}$ at an angular 
resolution, defined as the full width half maximum (FWHM) of the synthesised beam, of $\sim 6\arcsec$ across the whole northern hemisphere, using the High Band Antenna (HBA) system of LOFAR (see Fig.\ \ref{fig:lotss_summary}). 

In the first paper of this series (Paper I: \citealt{Shimwell_2017}) we described LoTSS and presented a preliminary
data release. In
that release the desired imaging specifications were not reached, as
no attempt was made to correct either for errors in the beam models or
for direction-dependent ionospheric distortions, which are severe in
these low-frequency data sets. However, there has since been substantial improvements in the quality, speed, and
robustness of the calibration of direction-dependent effects (DDEs)
and imaging with the derived solutions (see e.g.
\citealt{Tasse_2014b}, \citealt{Yatawatta_2015},
\citealt{vanWeeren_2016a} and \citealt{Tasse_2017}). Furthermore,
LOFAR surveys of smaller areas of sky have demonstrated that the
desired imaging specifications of LoTSS are feasible by making use of direction-dependent calibration (e.g. \citealt{Williams_2016} and \citealt{Hardcastle_2016}). These new insights have facilitated the first full quality public data release (LoTSS-DR1), which we present here in Paper II of this series.

As part of this series we also attempt to enrich our radio catalogues by
locating optical counterparts using a combination of likelihood ratio 
cross matching and visual inspection (discussed in Paper III of this
series: \citealt{Williams_2018}). In addition, where counterparts are
successfully located, we provide photometric redshift estimates and host galaxy properties (Paper
IV: \citealt{Duncan_2018c}). In the near future, to improve on the
redshifts for many sources, the William Herschel Telescope
Enhanced Area Velocity Explorer (WEAVE;
\citealt{Dalton_2012,Dalton_2014}) multi-object and integral field
spectrograph will measure redshifts of over a million LoTSS sources as part of the WEAVE-LOFAR survey (\citealt{Smith_2016}).

In Sec.\ \ref{sec:data_reduction} and \ref{sec:image_quality} we
describe the observations, the data processing procedure for the present data release, and the quality of the resulting images. In Sec.\ \ref{sec:cataloging} we give a brief overview of the optical cross matching and the photometric redshift estimation. Finally, we outline some upcoming developments in Sec.\ \ref{sec:futureprospects} before concluding in Sec.\  \ref{sec:conclusions}.

\begin{figure}   \centering
   \includegraphics[width=\linewidth]{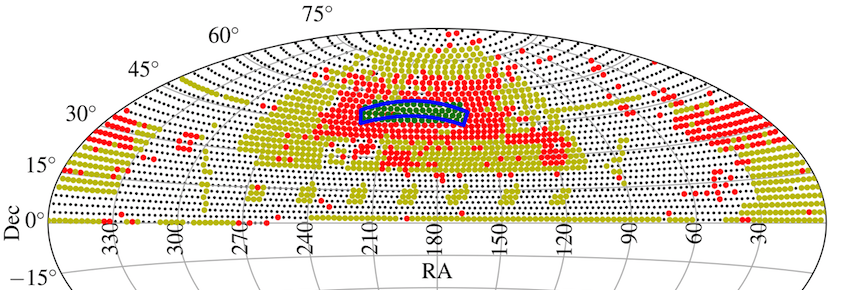}
   \caption{Status of the LoTSS observations as of May 2018. The green dots show the images that are presented in this paper. The red, yellow, and black dots show the observed pointings (but yet unpublished), pointings presently scheduled for observation between May 2018 and May 2020, and unobserved pointings, respectively. The HETDEX Spring Field region is outlined in blue. The vast majority of the completed coverage (20\% of the northern sky) and upcoming observations (an additional 30\% of the northern sky) are regions with low Galactic extinction.}
   \label{fig:lotss_progress}
\end{figure}

\section{Observations and data reduction}
\label{sec:data_reduction}

We describe the status of LoTSS observations in the first subsection. The second subsection outlines the direction-independent
calibration of the data;  at present, the main challenge is retrieving and processing the large volume of archived data. The third subsection describes the direction-dependent calibration and imaging, where the focus is on the development and execution of a robust and automated pipeline. The final subsection summarises the mosaicing and cataloguing of the DR1 images.

\subsection{Observation status}

The ambitious observational objectives for LoTSS are outlined in  Fig.\ \ref{fig:lotss_summary}. To achieve these objectives at optimal declinations, LoTSS observations are conducted in the \textsc{hba\_dual\_inner} configuration with 8\,hr dwell times and a frequency coverage of 120-168\,MHz. The entire northern sky is covered with 3168 pointings. By exploiting the multi-beam capability of LOFAR and observing in 8-bit mode two such pointings are observed simultaneously. As of May 2018, approximately 20\% of the data have now been gathered and a further 30\% are scheduled over the next two years (see Fig.\ \ref{fig:lotss_progress}); a total of approximately 13,000\,hr observing time are required to complete the entire survey with the present capabilities of LOFAR.

As in \cite{Shimwell_2017}, in this paper we focus on 63 LoTSS data sets (2\% of the total survey) in the region of the HETDEX Spring Field that were observed between 2014 May 23 and 2015 October 15. Each 8\,hr observation was bookended by 10\,min calibrator observations (primarily 3C 196 and 3C 295) and the data are archived with a time resolution of 1\,s and a frequency resolution of 16 channels per 195.3\,kHz sub-band (SB) by the observatory\footnote{$\sim$ 100 of the early LoTSS observations were averaged to 2\,s and
    24.4\,kHz}. This high time- and frequency-resolution data is kept to reduce time and bandwidth smearing to a level that is tolerable for future studies  that will exploit the international baselines of LOFAR (only antennas within the Netherlands are used for the primary objectives of LoTSS). The high spectral resolution (R$\sim$5000-7000 or 22-31\,km/s velocity resolution) of the data is also facilitating spectral line (\citealt{Emig_2018}) and spectro-polarimetric studies.

\subsection{Direction-independent calibration}

The publicly available LOFAR direction-independent calibration procedure was described in detail by \cite{vanWeeren_2016a} and \cite{Williams_2016} and makes use of the LOFAR Default Preprocessing Pipeline (DPPP; \citealt{vanDiepen_2018}) for averaging and calibration and BlackBoard Selfcal (BBS; \citealt{Pandey_2009}) for calibration. In Paper I we used a pipeline implementation\footnote{\url{https://github.com/lofar-astron/prefactor} using commit dd68c57} of this procedure to process the 63 LoTSS data sets that are described in this publication and we discussed the quality of the images that were produced. This calibration method is not described again in detail in this work, but we developed new tools to maintain a high volume flow of data through this pipeline and we briefly describe these below.

The LoTSS data are stored in the LOFAR Long Term Archive
(LTA{\footnote{\url{https://lta.lofar.eu/}}), which is distributed over
  three sites -- SURFsara\footnote{\url{https://www.surfsara.nl}},
  Forschungszentrum J\"{u}lich\footnote{\url{http://www.fz-juelich.de}}, and Pozna\'{n}\footnote{\url{http://www.man.poznan.pl/online/pl/}}. The archived data volume per 8\,hr
  pointing is $\sim$16\,TB, together with $\sim$350\,GB for each
  10\,min calibrator observation, which implies an eventual data volume of
  $\sim$50\,PB for the entire 3168 pointings of the survey (although this will be reduced by implementation of the DYSCO compression algorithm; \citealt{Offringa_2012}). Downloading these large data sets from the LTA sites to local facilities is either prohibitively time consuming or expensive.  To mitigate this we migrated our direction-independent calibration processing to the SURFsara Grid facilities. At the time of writing this consists of several hundred nodes of various sizes with a total of $\sim$7500 compute cores that are linked with a high-speed connection of 200\,Gbit/s peak network traffic to the Grid storage, where the SURFsara LTA data are housed. The implementation of the direction-independent calibration pipeline, and other LOFAR pipelines, on the SURFsara Grid is described in detail by \cite{Mechev_2017} and \cite{Oonk_2018} and summarised briefly below.

The LoTSS data are archived as 244 single SB files and in our SURFsara implementation of the direction-independent calibration pipeline each SB of the calibrator is sent to an available compute node where it is flagged for interference with \textsc{AOFLAGGER} (\citealt{Offringa_2012}), averaged to two channels per 195\,kHz SB and 8 s, and calibrated using a model of the appropriate calibrator source, which has a flux density scale consistent with that described in \cite{Scaife_2012}. We note that the \cite{Scaife_2012} flux density scale is consistent with the \cite{Perley_2017} scale to within $\sim$5\% but that there are larger discrepancies ($\sim$5-10\%) when comparing with the \cite{Baars_1977} scale (see \citealt{Scaife_2012} and \citealt{Perley_2017} for details). Using a single compute node the resulting 244 calibration tables are combined and used to derive time-independent amplitude solutions, XX and YY phase offsets, and clock offsets for each station. Similarly, on separate compute nodes, the 244 single SB target files are each flagged, corrected for ionospheric Faraday rotation\footnote{\url{https://github.com/lofar-astron/RMextract}}, calibrated using the calibrator solutions, and averaged to a resolution of two channels per 195\,kHz SB and 8 s. In the final step of the direction-independent calibration pipeline, the data for each contiguous 10-SB block are sent to different compute nodes where they are each combined to a single file that is phase calibrated against a sky model for the target field, which is generated from the TGSS-ADR1 catalogue (\citealt{Intema_2017}). This produces 25 10-SB measurement sets for the target field, but the six highest
frequency SBs are empty because there are only 244 SBs in the highest frequency measurement set. 

For the bulk processing of data on the SURFsara facilities we made use of PiCaS\footnote{\url{http://doc.grid.surfsara.nl/en/latest/Pages/Practices/picas/picas\_overview.html}}, a CouchDB based token pool server for heterogeneous compute environments. The PiCaS server allows millions of tasks to be scheduled on heterogeneous resources to monitor these tasks via a web interface and to provide easy access to logs and diagnostic plots, which helps ensure that our data quality is high. Examples of these diagnostic plots for the HETDEX Spring Field data are shown by \cite{Shimwell_2017}. We also make use of archiving and distribution facilities at SURFsara, allowing us to store the direction-independent calibrated data products (which are reduced from 16\,TB to $\sim$500\,GB per pointing) and freely distribute these amongst LoTSS team members for analysis and further processing.

The SURFsara Grid processing facilities enable high-throughput processing of large data sets stored on the local LTA site, however the LoTSS data sets are disseminated to all three LTA sites. Since the LTA sites are not linked to each other with a high bandwidth connection, the transfer speed to download data from the Forschungszentrum J\"{u}lich and Pozna\'{n} LTA sites to SURFSara ($\sim$200\,MB/s) is currently a bottleneck in our processing. We are therefore working on implementing the direction-independent calibration pipeline on compute facilities local to each of the LTA sites. 

\subsection{Direction-dependent calibration and imaging}

\begin{figure*}  \centering
   \includegraphics[width=0.43\linewidth]{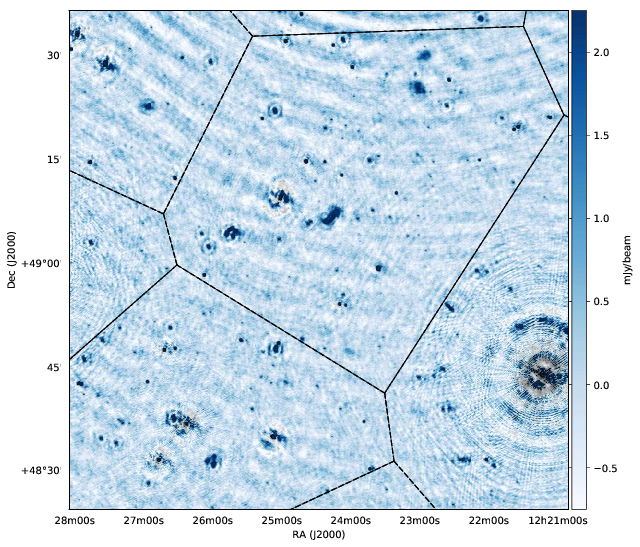}
   \includegraphics[width=0.43\linewidth]{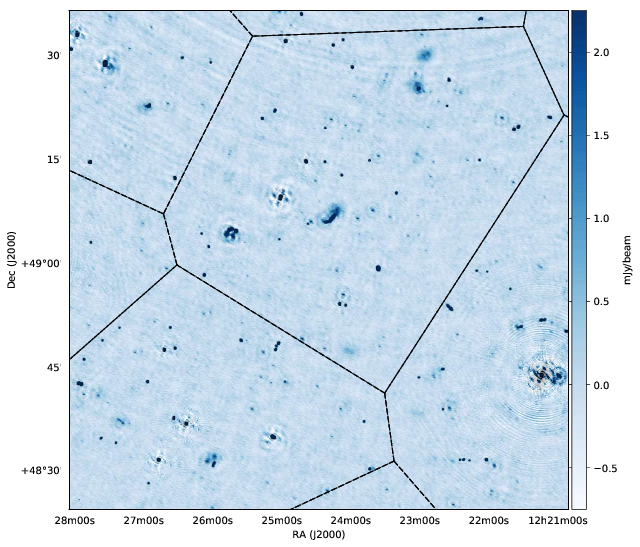} \\
   \includegraphics[width=0.43\linewidth]{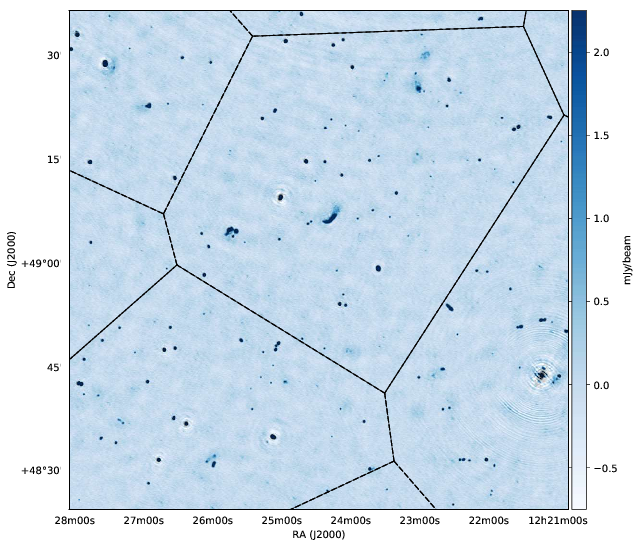}
   \includegraphics[width=0.43\linewidth]{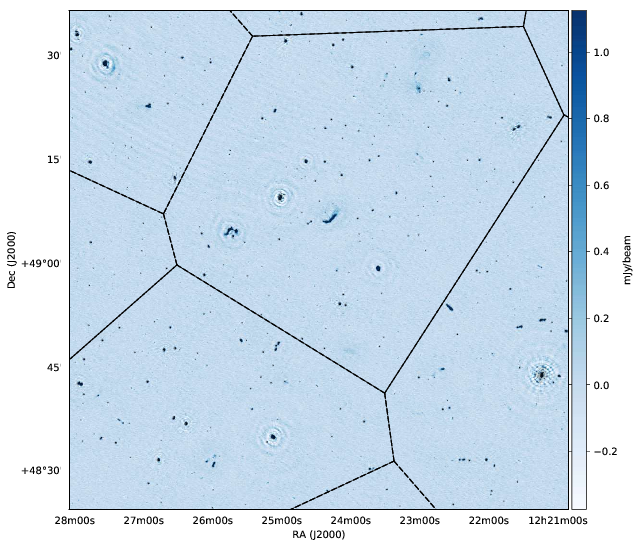}
   \caption{Self-calibration loop of LoTSS-DR1. From left to right top to bottom, the images show 60 SBs without any DDE correction, 60 SBs after applying DDE phase calibration, 60 SBs after applying DDE phase and amplitude calibration, and a 240 SB image after applying DDE phase and amplitude calibration. The colour scales are proportional to the square root of the number of SBs and the black lines show the facets used by \kMS/ and \DDF/.}
   \label{fig:ddfacet_selfcal}
\end{figure*}

A robust, fast, and accurate calibration and
imaging pipeline is essential to routinely create high-fidelity LoTSS images with a resolution of 6$\arcsec$ and a sensitivity of
100$\mu$Jy beam$^{-1}$. However the necessity to correct DDEs, which are primarily ionospheric distortions and errors in the station beam model of the HBA phased array stations, adds significant complications to this procedure.  These DDEs can be understood in terms of Jones matrices (\citealt{Hamaker_1996}) and to correct for these  matrices, which not only depend on direction but also on time, frequency, and antenna, they must be derived from the visibilities and applied during imaging. Various  approaches have been developed to estimate the DDE (e.g. \citealt{Cotton_2004}, \citealt{Intema_2009}, \citealt{Kazemi_2011}, \citealt{Noordam_2010},  \citealt{vanWeeren_2016a} and \citealt{Yatawatta_2015}) but for this work we developed KillMS (\kMS/; \citealt{Tasse_2014b} and \citealt{Smirnov_2015}) to calculate the Jones matrices and \DDF/ (\citealt{Tasse_2017}) to apply these during the imaging. Our software packages and the pipeline  are publicly available and documented\footnote{see \url{https://github.com/saopicc} for \kMS/ and \DDF/,
  and \url{https://github.com/mhardcastle/ddf-pipeline} for the
  associated \LOTTSpipe/ pipeline.}. Below we briefly outline the calibration and deconvolution procedures before describing the pipeline in more detail.

\subsubsection{Calibration of direction-dependent effects}
\label{sec:kMS}

One of the main
difficulties in the calibration of DDE is the large number of free parameters that must be optimised for 
when solving for the complex-valued Jones matrices.
The consequences of this are that finding the solutions can become
prohibitively computationally expensive and that ill-conditioning can
introduce systematics in the estimated quantities, which have a
negative impact on the image fidelity. 

To tackle the computational expense,  \cite{Salvini_2014}, \cite{Tasse_2014b}, and \cite{Smirnov_2015}  have shown
that when inverting the Radio Interferometeric Measurement
Equation (RIME; see e.g. \citealt{Hamaker_1996},  \citealt{Smirnov_2011}) the Jacobian
can be written using Wirtinger derivatives. The resulting Jacobian is remarkably
sparse, which allows for shortcuts to be used when implementing optimisation algorithms such as Levenberg-Marquardt (see for example Smirnov \& Tasse 2015). In particular, the problem can
become antenna separable, and to solve for the Jones matrices
associated with a given antenna in \kMS/, only the visibilities involving that
antenna are required at each iterative step. The computational gain
can be as high as $n_a^2$ (where $n_a$ is the number of elementary antennas).

To reduce ill conditioning, \kMS/ uses
the Wirtinger Jacobian together with an Extended Kalman Filter (EKF) to
solve for the Jones matrices (\citealt{Tasse_2018}). Instead of optimising
the least-squares residuals as a Levenberg-Marquardt (LM) procedure would,
the EKF is a minimum mean-square error estimator and is recursive
(as opposed to being iterative). In practice, the prior knowledge is used to constrain the expected solution
at a given time. While an LM would produce independent
``noisier" estimates, the EKF produces smooth solutions that are more physical and robust to ill-conditioning.  

To further improve the calibration, kMS produces a set of weights
according to a ``lucky imaging" technique in which the weights of
visibilities are based on the quality of their calibration solutions
(\citealt{Bonnassieux_2017}), so visibilities with the worst
ionospheric conditions are weighted down in the final imaging.

\subsubsection{Wide field spectral deconvolution}
\label{sec:DDF}

The \DDF/ imager (\citealt{Tasse_2017}) uses the \kMS/-estimated
direction-dependent Jones matrices and internally works on each of the
directions for which there are solutions to synthesise a single image.
To do this, several technical challenges had to be overcome. For
example, the dependence of the Jones matrices on time, frequency,
baseline, and direction, together with time- and frequency-dependent
smearing, lead to a position dependent point spread function (\PSF/).
Therefore, although \DDF/ synthesises a single image, each facet has
its own \PSF/ that takes into account the DDE and time and
bandwidth smearing whilst ensuring that the correct deconvolution problem is inverted in minor cycles. 

Furthermore, to accurately deconvolve the LoTSS images, which have a
large fractional bandwidth and a wide field of view, spectral
deconvolution algorithms must be used to estimate the flux density and
spectra of modelled sources whilst taking into account the variation of the LOFAR beam throughout the bandwidth of the data. The computational cost of this deconvolution can be high and therefore throughout our processing we make exclusive use of the subspace deconvolution (\SSD/) algorithm, an innovative feature of \DDF/ (see \citealt{Tasse_2017} for a description). As opposed to \CLEAN/ and related algorithms, where a fraction of the
flux density is iteratively removed at each major iteration, \SSD/ aims at
removing all the flux density at each major cycle. This is done in the
abstracted notion of subspaces --- in practice islands --- each representing
an independent deconvolution problem. Each one of
these individual subspaces is jointly deconvolved (all pixels are simultaneously
estimated) by using a genetic algorithm (the \SSDGA/
flavour of \SSD/), and parallelisation is done over hundreds
to thousands of islands.
A strength of
  \SSD/ is that we can minimise the number of major cycles, by always
  recycling the sky model from the previous 
  step. In practice the sky model generated in the preceding deconvolution step of the pipeline is then used to initialise the sky model in the next deconvolution. In other words, a proper dirty image is only formed at the very first imaging step and, thanks to \SSD/, the \LOTTSpipe/ pipeline can work only on residual images and update the spectral sky model at each deconvolution step.

\subsubsection{The \LOTTSpipe/ pipeline}
\label{sec:LOTTSpipeline}

The \LOTTSpipe/ pipeline has many configurable parameters including
resumability, taking into account time and bandwidth smearing,
bootstrapping the flux density scale off existing surveys, correction
of facet-based astrometric errors, user specified deconvolution
masks, and substantial flexibility in calibration and imaging
parameters. The pipeline is suitable for the analysis of various LOFAR
HBA continuum observations, including interleaved observations or
those spanning multiple observing sessions. The entire
pipeline takes less than five days to image one LoTSS pointing when
executed on a compute node with 512 GB RAM (the minimum required for
the pipeline is 192 GB) and four Intel Xeon E5-4620 v2 processors,
which have eight cores each (16 threads) and run at 2.6\,GHz.

The pipeline operates on the direction-independent calibration
products which, for each pointing, are 25 10-SB (1.95\,MHz)
measurement sets with a time and frequency resolution of 8\,s and two
channels per 195\,kHz SB. The pipeline first removes severely
flagged measurement sets (those with $\geq80\%$ of data flagged) and selects six 10-SB
blocks of data that are evenly spaced across the total bandwidth for
imaging. This quarter of the data is self-calibrated to gradually
build up a model of the radio emission in the field, which is then used
to calibrate the full data set. A brief outline of the steps of
\LOTTSpipe/, which are shown in Fig. \ref{fig:ddfacet_selfcal}, is as follows:

\begin{enumerate}[label=\textbf{Step.\arabic*},ref=\textbf{Step.\arabic*},leftmargin=1.2cm]
\item \label{pipe:DIDeconv} Direction-independent spectral
  deconvolution and imaging ($6\times10$ SB)
\item \label{pipe:Clustering} Sky model tesselation in 45
  directions
\item \label{pipe:DDCal0} Direction-dependent calibration ($6\times10$
  SB, \kMS/ with EKF);
\item \label{pipe:Boot} Bootstrapping the flux density scale\item \label{pipe:DDDeconv0} Direction-dependent spectral deconvolution and imaging
  ($6\times10$ SB, phase-only solutions, three major cycles)\item \label{pipe:DDCal1} Direction-dependent calibration ($6\times10$ SB, \kMS/ with EKF)
\item \label{pipe:DDDeconv1} Direction-dependent spectral deconvolution and imaging
  ($6\times10$ SB), one major cycle, amplitude, and phase solutions)
\item \label{pipe:DDCal2} Direction-dependent calibration ($24\times10$ SB, \kMS/ with EKF)
\item \label{pipe:DDDeconv2} direction-dependent spectral deconvolution and imaging
  ($24\times10$ SB, two major cycles, amplitude, and phase solutions)
\item \label{pipe:AstroCorr} Facet-based astrometric correction.
\end{enumerate}

The \DDF/ is used in \ref{pipe:DIDeconv} to image the direction-independent calibrated
data using the \SSD/ algorithm, which allows us
to rapidly deconvolve very large images. The present implementation of
\SSD/  requires a deconvolution mask and we use \DDF/ to
automatically generate one based on a threshold of 15 times the local
noise, which is re-evaluated at every major cycle. The mask created
during the deconvolution is supplemented with a mask
generated from the TGSS-ADR1 catalogue to ensure
that all bright sources in the field are deconvolved even when
observing conditions are poor and automatically masking the sources is
challenging. The image produced from the 60 SB data set consists
of 20,000$\times$20,000 1.5$\arcsec$ pixels, has a restoring beam of
$12\arcsec$, and the noise varies between 0.25\,mJy beam$^{-1}$ and
2\,mJy beam$^{-1}$ depending on the observing conditions and source
environment. From this image a refined deconvolution mask is created
and used to reduce the number of spurious components in the
\SSD/ component model of the field by filtering out those
that lie outside the region within the refined mask. 

At \ref{pipe:Clustering} the resulting sky model is
used to define 45 facets that cover the full
$8.3^{\circ}\times8.3^{\circ}$ region that has been imaged. The
\SSD/ component model is used for the first direction
dependent calibration of the 60 SB data set (\ref{pipe:DDCal0}). This calibration is
done using \kMS/, which creates an amplitude and phase
solution for each of the 45 facets every 60\,s and 1.95\,MHz of
bandwidth, and the data are reimaged. Throughout the pipeline, in order not to absorb unmodelled sky emission
  into the \kMS/ calibration solutions (in particular faint extended
  emission seen by a small number of baselines), we always calibrate
the visibilities using only baselines longer than $1.5$ km (corresponding to
scales of $\sim4.5\arcmin$).

After this initial direction-dependent calibration we bootstrap the
LoTSS-DR1 flux densities in \ref{pipe:Boot} following the procedure described by
\cite{Hardcastle_2016}. This not only improves the accuracy of our
flux density estimates but also decreases amplitude errors that can occur owing
to imperfections in the calibration across the bandwidth. In this step
each of the six 10-SB blocks imaged in the previous step is
imaged separately at lower resolution (20$\arcsec$) using \DDF/ which applies the direction
dependent phase calibration solutions. A catalogue is made from the resulting image cube
using the Python Blob Detector and Source Finder (PyBDSF;
\citealt{Mohan_2015}) where sources are identified using a combined
image created from all the planes in the cube and the source flux density 
measurements are extracted from each plane using the same
aperture. Sources within 2.5$^\circ$ of the pointing centre that are
at least 100$\arcsec$ from any other detected source and have a integrated
flux density exceeding 0.15\,Jy are positionally cross matched with the VLSSr and WENSS catalogues using matching radii of
40$\arcsec$ and 10$\arcsec$, respectively. The WENSS catalogue used has
all the flux densities scaled by a factor of 0.9 which, as described by \cite{Scaife_2012}, brings it into overall agreement with the flux density scale we use. Correction factors are then derived for each of the six 10-SB blocks to best align the LoTSS-DR1 integrated flux density measurements with VLSSr and WENSS assuming the sources have power-law profiles across this frequency range (74\,MHz to 325\,MHz). During the fitting, sources that are poorly described by a power law are excluded to remove, for example incorrect matches or sources with spectral curvature. From the 70$\pm$14 matched sources per field the correction factors derived for each of the six 10-SB blocks are typically 0.85$\pm$0.1 and these are extrapolated linearly to the entire 25 10-SB
data set.  The six 10-SB 20$\arcsec$ resolution images are also
stacked to provide a lower resolution (20$\arcsec$) image that has
a higher surface brightness sensitivity than the higher resolution
images. This image is used to identify diffuse structures  that are prevalent in LOFAR images, but may not be 
detected at sufficient significance in the higher resolution
imaging. These extended sources are then added to the mask to
ensure that they are deconvolved in later imaging steps. Sources are
classified as extended sources if they encompass a contiguous region
larger than 2000 pixels with all pixels having a signal above three times the local
noise of the image.

After the bootstrap derived corrections factors are applied the 60
SBs of data are imaged with the direction-dependent phase
solutions applied in \DDF/ in \ref{pipe:DDDeconv0}. As explained
above, for efficiency reasons \SSD/  is
initiated with the \SSD/  components from the direction-independent imaging, which allows us to deconvolve deeply with three
major \SSD/ iterations. The image size and resolution are the
same as \textbf{in \ref{pipe:DDCal0}} but the input mask is improved because it is a
combination of that obtained from the direction-independent
imaging, the mask generated from the TGSS-ADR1 catalogue, and the
low-resolution mask created from the bootstrapping; at this point the auto-masking threshold is also lowered to ten times the local noise. Again,
once the imaging is complete the image is masked and the mask is used
to reduce spurious entries in the \SSD/ component model. The
noise levels in this second imaging step range from 130$\mu$Jy beam$^{-1}$ to
600$\mu$Jy beam$^{-1}$. In \ref{pipe:DDCal1} this new model is input into \kMS/ which
calculates improved direction-dependent calibration solutions for each
of the 45 facets every 60\,s and 1.95\,MHz of bandwidth.

A third imaging step is performed on the 60 SBs of data (\ref{pipe:DDDeconv1}), this
time applying both the phase and amplitude direction-dependent
calibration solutions but otherwise following the same procedure as
before. This produces images with noise levels ranging from
100$\mu$Jy beam$^{-1}$ to 500$\mu$Jy beam$^{-1}$ and a final \SSD/
component model that is used to calibrate the entire 240 SBs of
the data set with \kMS/ (\ref{pipe:DDCal2}).

The full bandwidth is imaged at both low and high resolution in
\DDF/ with the newly derived phase and amplitude solutions
applied (\ref{pipe:DDDeconv2}). The low-resolution image has a resolution of 20$\arcsec$ and
a significantly higher surface brightness sensitivity than when
imaging at higher resolution. In this low-resolution image \SSD/ is not initiated with a
previously derived model because the $uv$-data used in the imaging are
different as an outer $uv$-cut of 25.75\,km is applied. To deconvolve deeply we perform three separate
iterations of the low-resolution imaging, each time improving the input
mask and lowering the automasking threshold. The noise level of the
final 20$\arcsec$ resolution images ranges from 100$\mu$Jy beam$^{-1}$ to
400$\mu$Jy beam$^{-1}$, which corresponds to a brightness temperature of 9\,K
to 35\,K.

The full bandwidth high-resolution imaging is performed with a
resolution of 6$\arcsec$. The deconvolution mask that has been gradually
built up through the self-calibration of the 60 SB data set, as
well as that from the lower resolution imaging from the full
bandwidth, and an auto masking threshold in \DDF/ of five times the local
noise allow
for a very deep deconvolution. This is performed with two separate
runs of \DDF/ with a masking step in between to ensure that the local noise is well estimated and faint sources (signal to noise $\geq$5) are masked. The resulting high-resolution images have noise
levels that vary from 60$\mu$Jy beam$^{-1}$ to 160$\mu$Jy beam$^{-1}$. Once the
deconvolution is complete the images are corrected for astrometric
errors in \DDF/ which can apply astrometric corrections to
each of the facets independently (\ref{pipe:AstroCorr}). The astrometric corrections applied
vary from 0.0$\arcsec$ to 4.4$\arcsec$ with a median of 0.8$\arcsec$
and are derived from cross-matching the LOFAR detected sources in each
facet with the Pan-STARRS catalogue (\citealt{Flewelling_2016}). The errors on the derived offsets vary from 0.1$\arcsec$ to 4.8$\arcsec$ with a median 0.2$\arcsec$.

During the cross-matching a histogram of the separations between all
Pan-STARRS sources within 60 arcsec of compact LOFAR sources is made
for each facet. This typically consists of $\sim$140 Pan-STARRS
sources per LoTSS-DR1 source and an average of 190 radio sources per
facet. If all sources in the facets are systematically offset, then
this histogram should have a peak at the value of the offset between
the LoTSS-DR1 and Pan-STARRS sources. To search for the location of this
peak and estimate the RA and Dec offsets and their corresponding
errors in each facet, we use a Markov Chain Monte Carlo (MCMC) method
and uninformative priors. In this procedure the \textit{emcee} package (\citealt{Foreman_2013}) is used to draw MCMC samples from a Gaussian function plus a
background where the initial parameter estimates are derived from the
observed LOFAR and Pan-STARRS position offsets. The likelihood function
is calculated using a gamma distribution with a shape parameter
defined by the observed LOFAR and Pan-STARRS position offsets. The
posterior probability distribution is calculated taking into account
the uninformative priors (background, offset, and Gaussian peak greater
than zero and a Gaussian standard deviation less than 5$\arcsec$) that
are put on the offset Gaussian function and background level.

The pipeline is very robust and with no human interaction the
processing failed for only 5 of the 63 fields in the HETDEX Spring
Field region, thus providing 58 images in this region. One (P2) of
these failures was due to exceptionally bad ionospheric conditions and
the other four (P31, P210+52, P214+52, and P215+50) were due to the
proximity of very bright sources (3C 280 and 3C 295). 

\subsection{Mosaicing and radio source cataloguing}
\label{sec:mosaicing} 

\begin{figure*}   \centering
   \includegraphics[width=\linewidth]{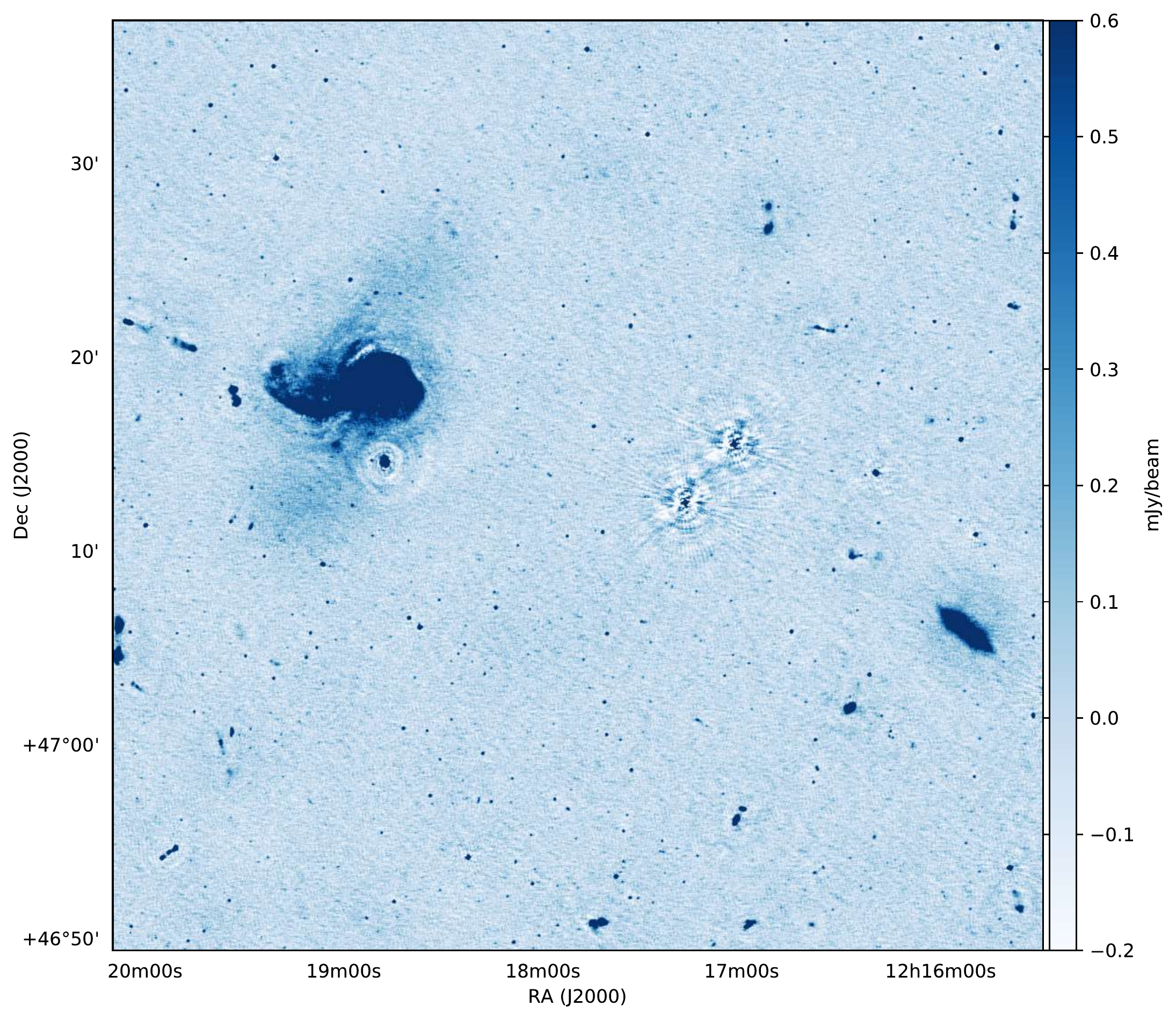}
   \includegraphics[width=0.45\linewidth]{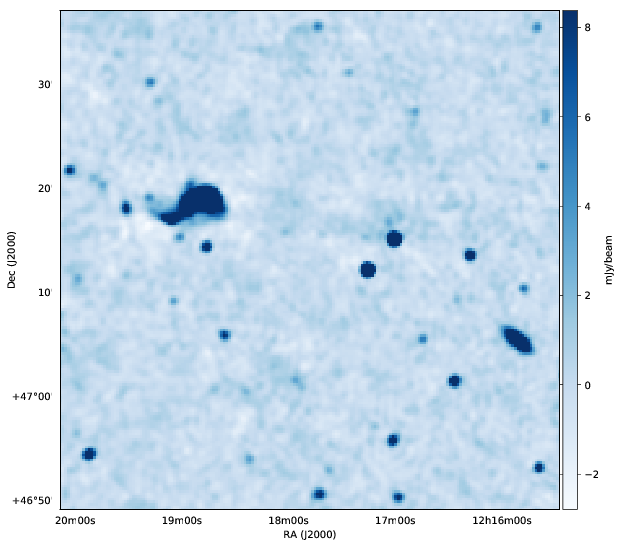}
   \includegraphics[width=0.45\linewidth]{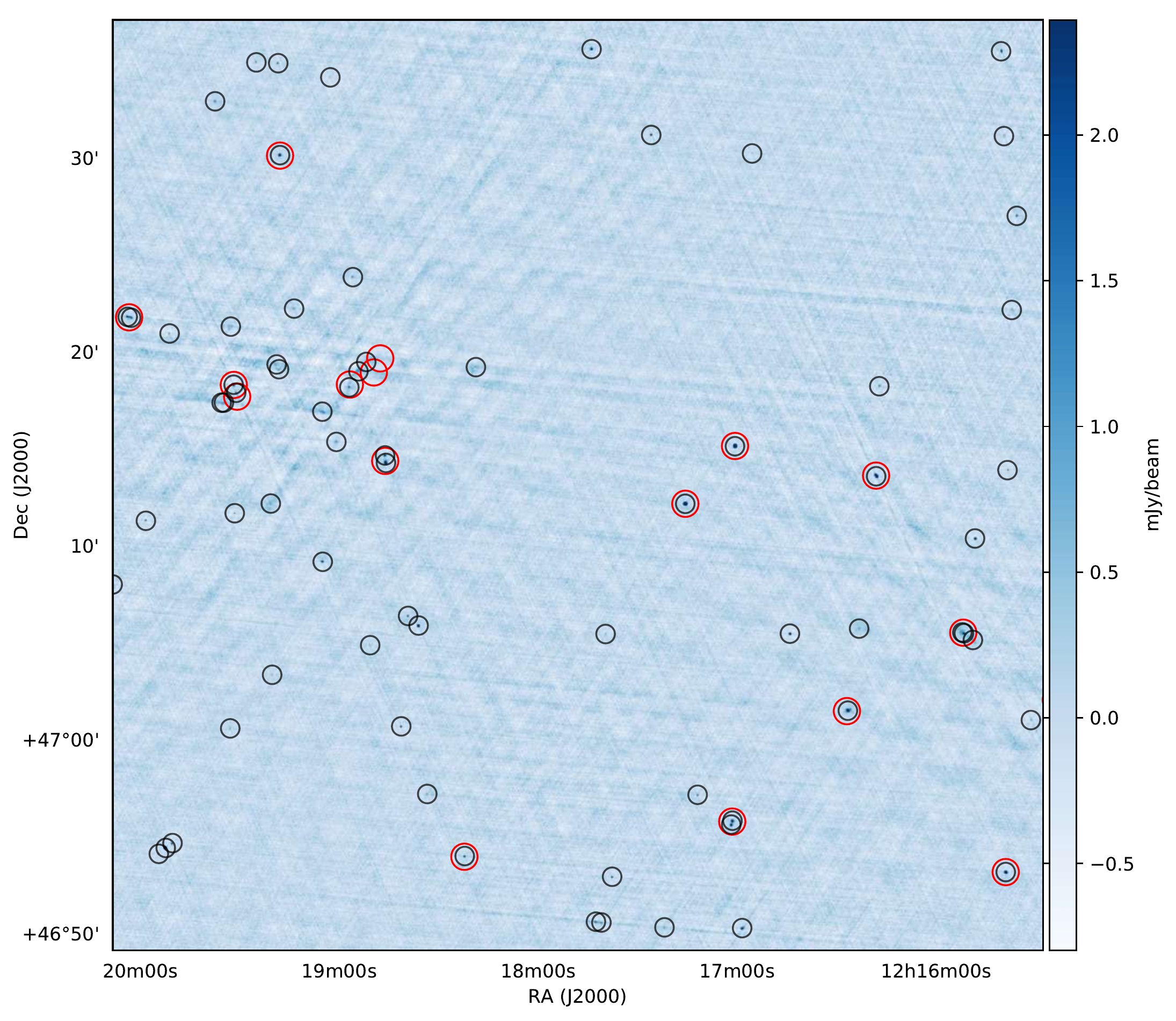}
   \caption{Top figure shows an example of a LoTSS-DR1 image and bottom figures show the same region in NVSS (left) and FIRST (right). The black and red circles overlaid on the FIRST image show FIRST and TGSS-ADR1 sources, respectively. 
  In this region there are 689 LoTSS-DR1 sources, 71 FIRST sources,
  46 NVSS sources, and 16 TGSS-ADR1 sources. The resolution of the
  LoTSS-DR1 image is 6$\arcsec$ and the sensitivity in this region is
  approximately 70$\mu$Jy beam$^{-1}$. This field is dominated by the
  spectacular galaxy NGC 4258, which in the LoTSS-DR1 image has an
  extent of over 3000 synthesised beams, together with the smaller edge-on spiral galaxy NGC 4217.}
   \label{fig:mosaic-example} 
\end{figure*}

\begin{figure*}   \centering
   \includegraphics[width=1.0\linewidth]{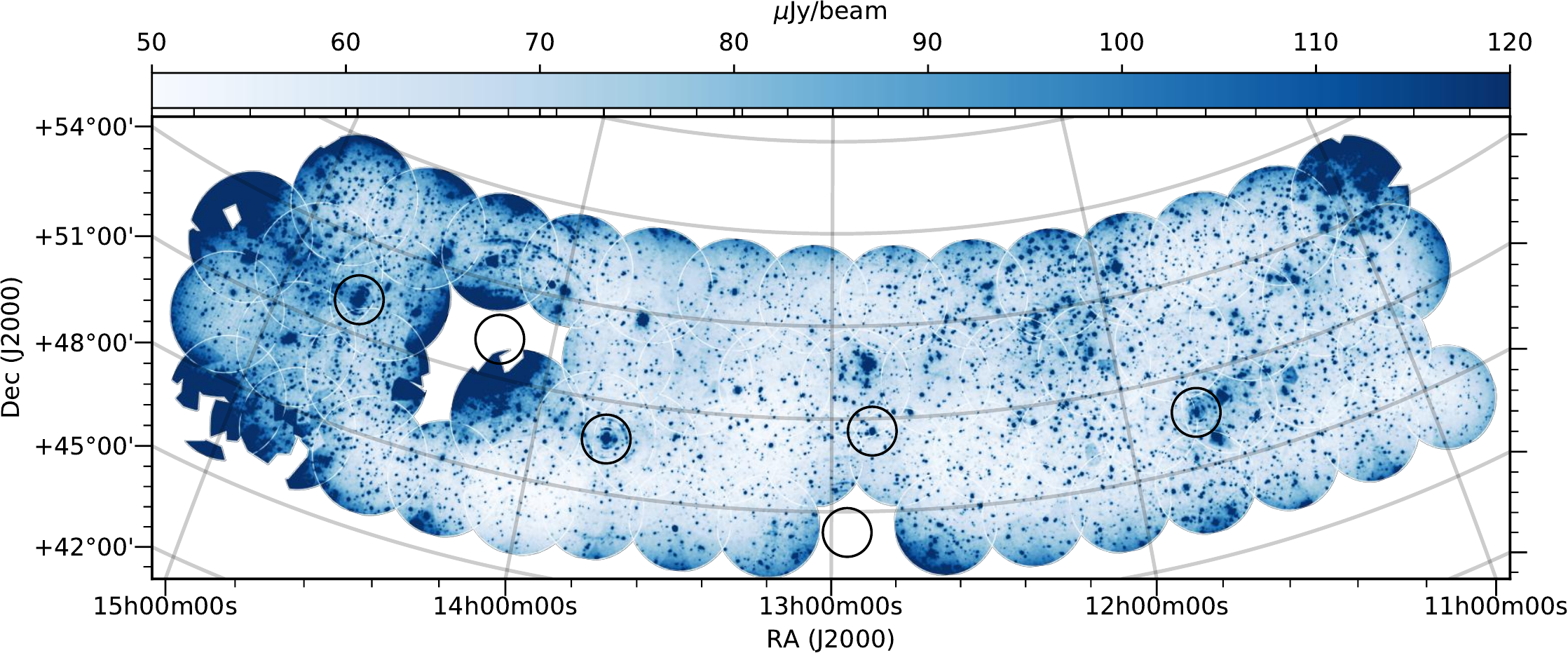}
   \caption{Noise image of the LoTSS-DR1 where the median noise
     level is 71$\mu$Jy beam$^{-1}$. As described in
     Sec.\ \ref{sec:dynamicrange} many of the regions with high noise levels
     are caused by dynamic-range limitations. Sources from the revised 3C catalogue
     of radio sources (\citealt{Bennett_1962}) are overplotted as black circles to show
     the location of potentially problematic objects.}
   \label{fig:mosaic-noisemap} 
\end{figure*}

\begin{figure*}   \centering
   \includegraphics[width=0.43\linewidth]{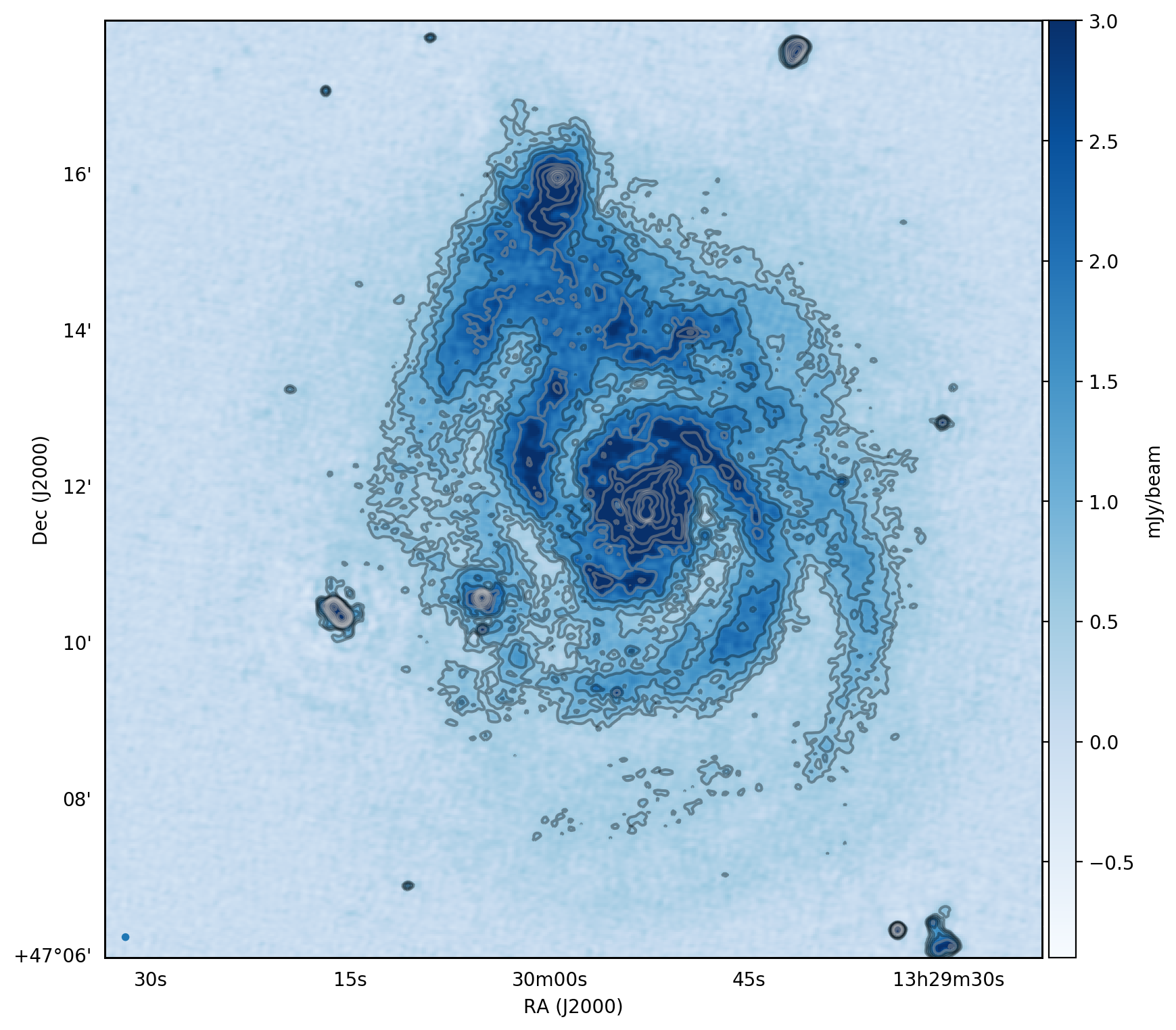}
   \includegraphics[width=0.43\linewidth]{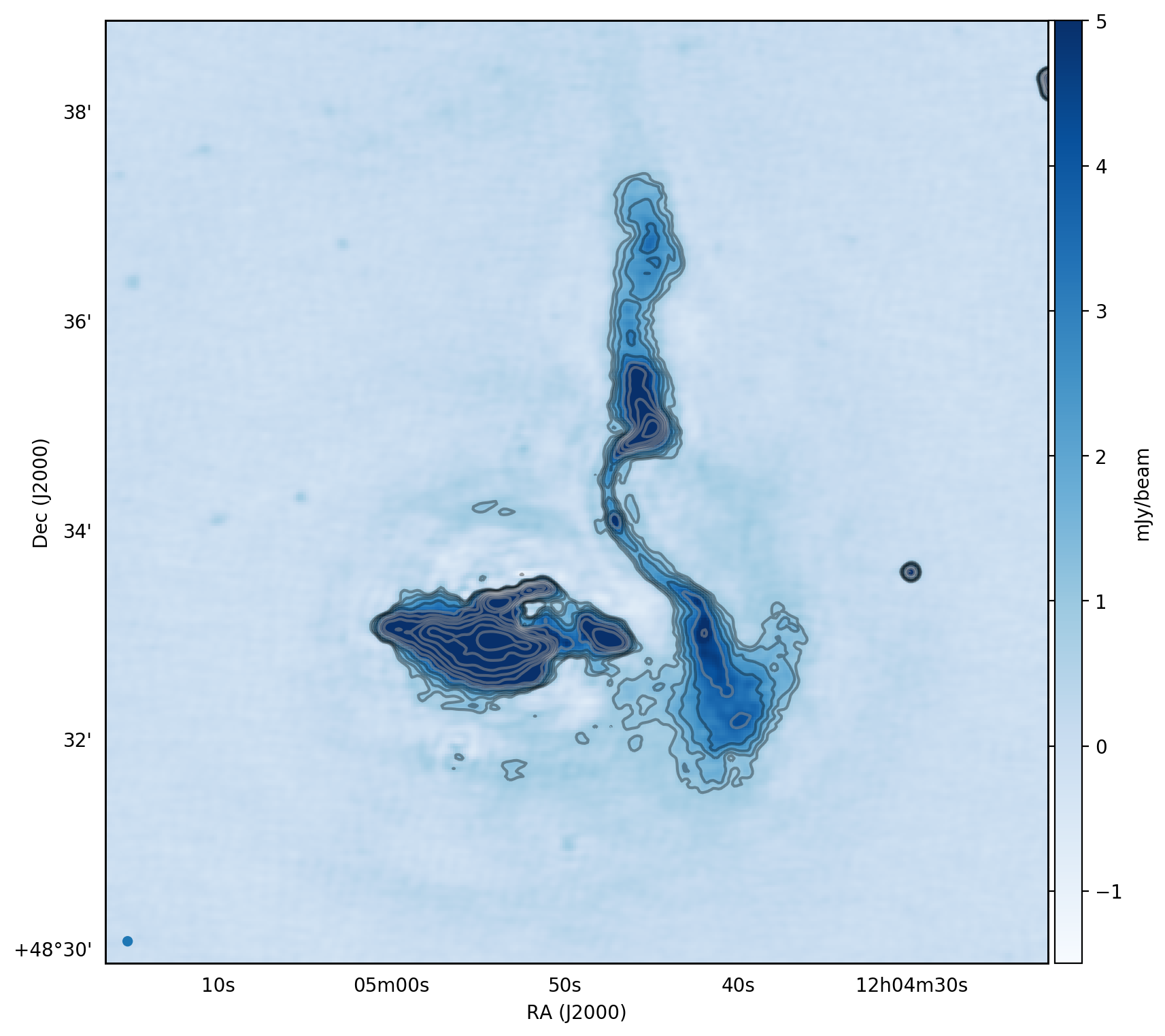}\\
   \includegraphics[width=0.43\linewidth]{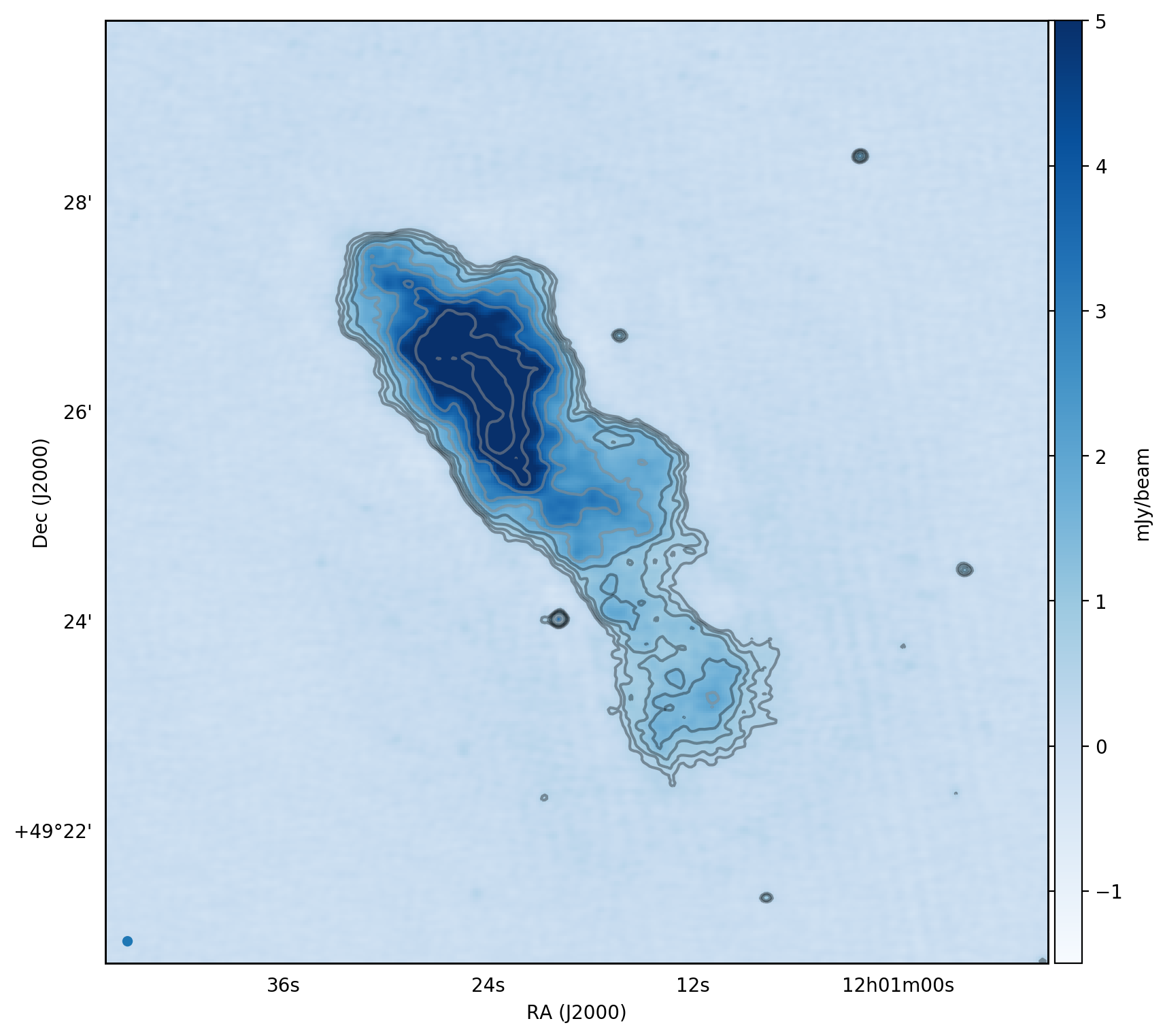}
   \includegraphics[width=0.43\linewidth]{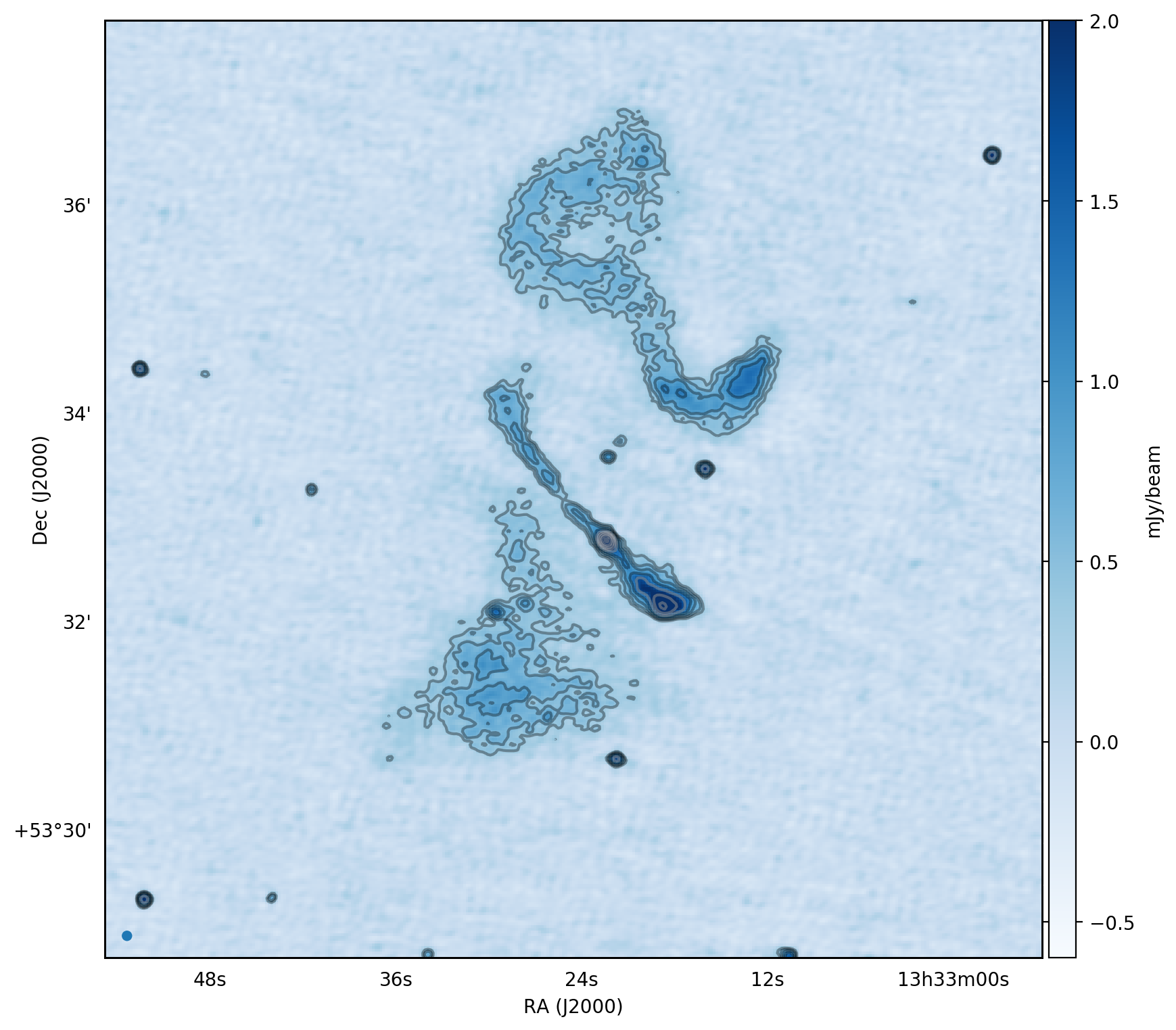}\\
   \includegraphics[width=0.43\linewidth]{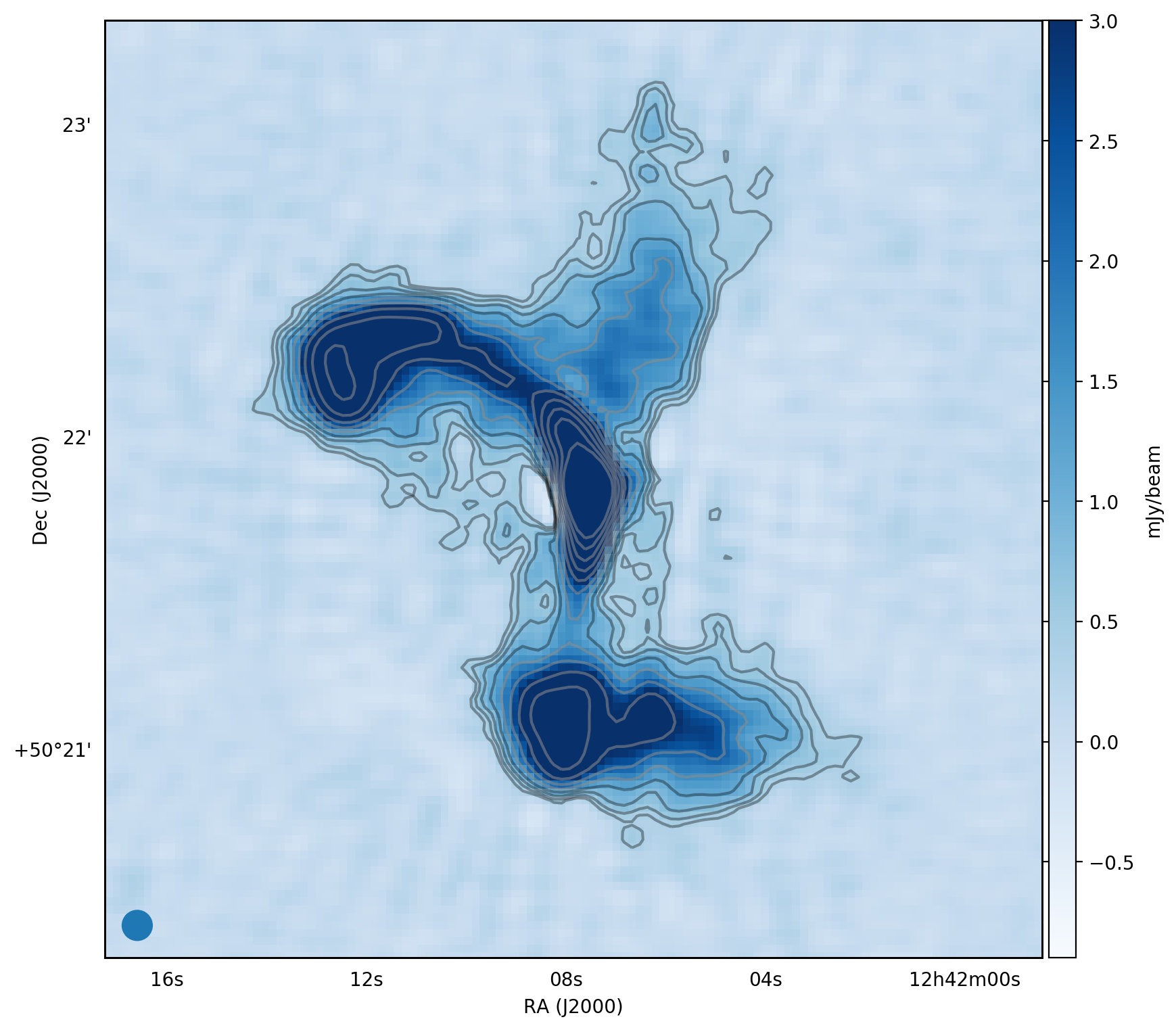}
   \includegraphics[width=0.43\linewidth]{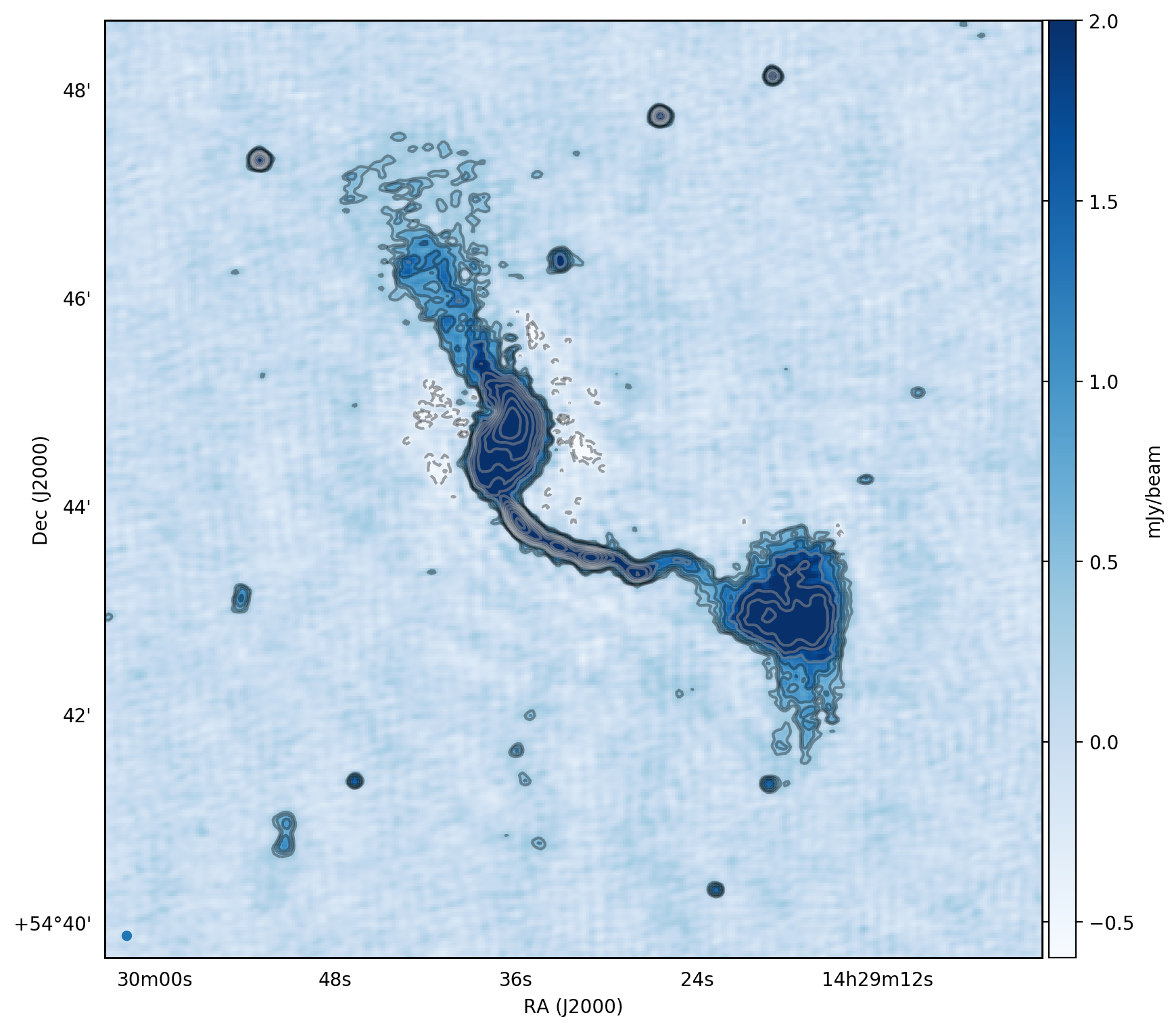}\\   
   \caption{Selection of resolved sources in the LoTSS-DR1 images with the colour scale and contours chosen for display purposes. The synthesised beam is shown in the bottom left corner of each image.}
   \label{fig:example-complex-sources} 
\end{figure*}

The LoTSS pointings tile the sky following a spherical spiral
distribution (\citealt{Saff_1997}); they are typically separated by
2.58$^\circ$ and have six nearest neighbours within 2.8$^\circ$. With
the FWHM of the \textsc{hba\_dual\_inner}
station beam being 3.40$^\circ$ and 4.75$^\circ$ at the top (168\,MHz) and bottom (120\,MHz) 
of the LoTSS frequency coverage, respectively, there is significant
overlap between the pointings. To produce the final data release
images, a mosaic  has been generated for each of the 58 pointings that was
successfully processed. For each pointing the images of the (up to
six) neighbouring pointings are reprojected to the frame of the
central pointing using the {\sc astropy}-based {\it reproject} code and then
all seven (or fewer) pointings are averaged using the appropriate
station beam and the central image noise as weights in the averaging.
During the mosaicing of the high-resolution images, facets with
uncertainties in the applied astrometric corrections  
(derived as described in Sec.~\ref{sec:LOTTSpipeline}) 
larger than 0.5$\arcsec$ are excluded to
ensure that the final maps have a high astrometric accuracy. This criterion is
also a good proxy for image quality and allows us to identify and
remove any facets that diverged during the processing due to poor
calibration solutions. Once the images of the neighbouring pointings
are combined the mosaiced map is blanked to leave just the pixels that
lay within the 0.3 power point of the station beam of the central
pointing. An example region from a mosaic is shown in
Fig.\ \ref{fig:mosaic-example} and the noise map of the entire
mosaiced region is shown in Fig.\ \ref{fig:mosaic-noisemap}.

To produce a catalogue of the radio sources we performed source detection on each mosaic using PyBDSF. The sources were detected with a 5$\sigma$ peak detection threshold and a 4$\sigma$ threshold to define the boundaries of the detected source islands that were used for fitting. The background noise variations were estimated across the images using a sliding box algorithm, where a box size of $30\times30$ synthesised beams was used except in the regions of high signal-to-noise sources ($\geq$150) where the box size was decreased to just 12$\times$12 synthesised beams; this box size was tuned to more accurately capture the increased noise level in these regions. The PyBDSF wavelet decomposition functionality was also utilised to better characterise the complex extended emission in the images. The resulting catalogues of the individual mosaics were combined and duplications were removed by only keeping sources that are detected in the mosaic to which they are closest to the centre.

In the concatenated catalogue the columns kept from the PyBDSF output
are the source position, peak brightness,  integrated flux density,
source size and orientation, and the statistical errors from the source fitting for each of these. In addition we keep the source code which describes the type of structure fitted by PyBDSF (see Table~1 caption for the definition of these) and the local root mean square noise estimate. We append columns that provide the mosaic identity, number of pointings that contribute to the mosaic at the position of the source, fraction of those in which the source was in the deconvolution mask, and whether or not the source is believed to be an artefact (see \citealt{Williams_2018} for a description of artefact identification). The fraction of the source in the deconvolution mask is calculated by finding the mask value (1 or 0) at the centre of each Gaussian component for every source in all of the contributing pointings and using the effective integration times to calculate the weighted average. To find the masked fraction for a source that consists of multiple Gaussian components, we use the integrated flux densities of each component as weights and assign the weighted average of the masked fraction of these components to the source. These final parameters, together with the mosaiced residual images, which are also provided, allow users to assess the quality of the deconvolution for sources. This is particularly important for faint sources that may not be in the masks and also for extended sources where, because of the integral of the dirty beam exceeding that of the restoring beam, the apparent flux density in dirty images is substantially larger than in deconvolved images. Example entries from the catalogue are shown in Table 1 and a selection of some of the more spectacular sources in our images are represented in Fig.\ \ref{fig:example-complex-sources}. 

\begin{sidewaystable*}
\centering
{{Table 1:} Example of entries in the LoTSS-DR1 source catalogue. The entire catalogue contains 325,694 sources. The entries in the catalogue are as follows: source identifier (ID), J2000 right ascension (RA), J2000 declination (Dec), peak brightness ($\rm{S_{peak}}$), integrated flux density ($\rm{S_{int}}$), major axis (Maj), minor axis (Min), deconvolved major axis (DC Maj), deconvolved minor axis (DC Min), position angle (PA),   local noise at the position of the source (RMS), type of source as classified by PyBDSF (Type -- where `S' indicates an isolated source that is fit with a single Gaussian; `C' represents sources that are fit by a single Gaussian but are within an island of emission that also contains other sources; and `M' is used for sources that are extended and fitted with multiple Gaussians), the mosaic identifier (Mosaic), the number of pointings that are mosaiced at the position of the source (Number pointings), the fraction of pointings in which the source is in the deconvolution mask (Masked fraction), and  whether or not the entry is believed to be an artefact (Artefact). Only 2590 entries have been identified as artefacts (see \citealt{Williams_2018}). The errors in the catalogue are the uncertainties obtained from the PyBDSF source fitting. Additional uncertainties on the source extensions, astrometry, and flux scale are described in Section \ref{sec:image_quality}.  }\\
\label{tab:catalogue-example}
\begin{tabular}{lccccccccccccccc}
\hline
Source ID & RA  & DEC  & $S_{peak}$  & $S_{int}$ & Maj & Min & DC\_Maj    & DC\_Min  & PA & RMS  & Type & Mosaic & Number & Masked & Arte-\\  
 &   &   & (mJy/  & (mJy) & ($\arcsec$) & ($\arcsec$)  &  ($\arcsec$)   &  ($\arcsec$)& ($^\circ$)  & (mJy/  &  & & pointings & fraction & fact \\ 
& & & (beam) &   & & & & & & (beam) \\  \hline
ILTJ140629.07 & 211.6211$^\circ$ & 54.0190$^\circ$ & 0.3 & 0.6  & 9.0 & 7.3  & 6.7  & 4.1  & 89 & 0.08 & S & P209+55 & 2 & 0.00 & $\times$ \\
+540108.3 & $\pm$1.2$\arcsec$ & $\pm$0.8$\arcsec$ & $\pm$0.1 & $\pm$0.1 & $\pm$2.9 & $\pm$2.0 & $\pm$2.9 & $\pm$2.0 & $\pm$64 \\
ILTJ113833.90 & 174.6413$^\circ$ & 52.4434$^\circ$ & 0.4 & 0.5  & 7.7 & 5.2  & 0.0  & 0.0  & 87 & 0.06 & S & P12Hetdex11 & 4 & 0.00 & $\times$ \\
+522636.3 & $\pm$0.6$\arcsec$ & $\pm$0.3$\arcsec$ & $\pm$0.1 & $\pm$0.1 & $\pm$1.3 & $\pm$0.6 & $\pm$1.3 & $\pm$0.6 & $\pm$19 \\
ILTJ114532.75 & 176.3864$^\circ$ & 47.9582$^\circ$ & 0.4 & 0.4  & 7.3 & 5.8  & 4.2  & 0.0  & 61 & 0.06 & S & P15Hetdex13 & 4 & 0.00 & $\times$ \\
+475729.5 & $\pm$0.6$\arcsec$ & $\pm$0.5$\arcsec$ & $\pm$0.1 & $\pm$0.1 & $\pm$1.5 & $\pm$0.9 & $\pm$1.5 & $\pm$0.9 & $\pm$36 \\
ILTJ114443.86 & 176.1827$^\circ$ & 46.5350$^\circ$ & 1.4 & 2.2  & 8.7 & 6.2  & 6.3  & 1.7  & 168 & 0.11 & S & P15Hetdex13 & 2 & 0.77 & $\times$ \\
+463206.0 & $\pm$0.2$\arcsec$ & $\pm$0.4$\arcsec$ & $\pm$0.1 & $\pm$0.2 & $\pm$0.8 & $\pm$0.5 & $\pm$0.8 & $\pm$0.5 & $\pm$12 \\
ILTJ134204.70 & 205.5196$^\circ$ & 49.5226$^\circ$ & 1.0 & 1.1  & 6.8 & 6.0  & 0.0  & 0.0  & 87 & 0.06 & S & P42Hetdex07 & 3 & 1.00 & $\times$ \\
+493121.4 & $\pm$0.2$\arcsec$ & $\pm$0.2$\arcsec$ & $\pm$0.1 & $\pm$0.1 & $\pm$0.5 & $\pm$0.4 & $\pm$0.5 & $\pm$0.4 & $\pm$21 \\
ILTJ111452.35 & 168.7181$^\circ$ & 54.7318$^\circ$ & 0.5 & 0.4  & 5.9 & 5.4  & 0.0  & 0.0  & 121 & 0.08 & S & P169+55 & 3 & 0.71 & $\times$ \\
+544354.6 & $\pm$0.4$\arcsec$ & $\pm$0.3$\arcsec$ & $\pm$0.1 & $\pm$0.1 & $\pm$0.9 & $\pm$0.8 & $\pm$0.9 & $\pm$0.8 & $\pm$70 \\
ILTJ122721.08 & 186.8378$^\circ$ & 50.8748$^\circ$ & 0.6 & 3.0  & 17.3 & 10.2  & 16.2  & 8.3  & 18 & 0.06 & S & P26Hetdex03 & 4 & 0.60 & $\times$ \\
+505229.1 & $\pm$0.4$\arcsec$ & $\pm$0.7$\arcsec$ & $\pm$0.1 & $\pm$0.1 & $\pm$1.8 & $\pm$0.9 & $\pm$1.8 & $\pm$0.9 & $\pm$10 \\
ILTJ121124.15 & 182.8506$^\circ$ & 47.0435$^\circ$ & 1.9 & 8.0  & 27.5 & 5.5  & 0.0  & 0.0  & 177 & 0.11 & M & P19Hetdex17 & 3 & 0.93 & $\times$ \\
+470236.5 & $\pm$1.8$\arcsec$ & $\pm$0.2$\arcsec$ & $\pm$0.1 & $\pm$0.3 & $\pm$4.2 & $\pm$0.4 & $\pm$4.2 & $\pm$0.4 & $\pm$6 \\
ILTJ143026.91 & 217.6121$^\circ$ & 48.0969$^\circ$ & 0.5 & 0.6  & 7.3 & 6.2  & 4.2  & 1.7  & 135 & 0.06 & S & P217+47 & 3 & 0.70 & $\times$ \\
+480548.8 & $\pm$0.4$\arcsec$ & $\pm$0.4$\arcsec$ & $\pm$0.1 & $\pm$0.1 & $\pm$1.0 & $\pm$0.7 & $\pm$1.0 & $\pm$0.7 & $\pm$35 \\
ILTJ124531.14 & 191.3797$^\circ$ & 53.6874$^\circ$ & 0.4 & 0.7  & 10.6 & 6.2  & 8.8  & 1.7  & 46 & 0.06 & S & P191+55 & 4 & 0.59 & $\times$ \\
+534114.7 & $\pm$0.7$\arcsec$ & $\pm$0.7$\arcsec$ & $\pm$0.1 & $\pm$0.1 & $\pm$2.1 & $\pm$0.8 & $\pm$2.1 & $\pm$0.8 & $\pm$17 \\
ILTJ144630.58 & 221.6274$^\circ$ & 52.8271$^\circ$ & 0.6 & 7.4  & 25.7 & 17.0  & 25.0  & 15.9  & 61 & 0.30 & S & P223+52 & 3 & 1.00 & $\surd$ \\
+524937.5 & $\pm$3.6$\arcsec$ & $\pm$2.8$\arcsec$ & $\pm$0.2 & $\pm$0.2 & $\pm$9.2 & $\pm$5.7 & $\pm$9.2 & $\pm$5.7 & $\pm$47 \\
ILTJ115435.71 & 178.6488$^\circ$ & 49.2769$^\circ$ & 0.3 & 0.4  & 9.9 & 5.0  & 0.0  & 0.0  & 102 & 0.06 & S & P18Hetdex03 & 5 & 0.00 & $\times$ \\
+491636.8 & $\pm$1.0$\arcsec$ & $\pm$0.4$\arcsec$ & $\pm$0.1 & $\pm$0.1 & $\pm$2.5 & $\pm$0.7 & $\pm$2.5 & $\pm$0.7 & $\pm$15 \\
ILTJ134314.24 & 205.8093$^\circ$ & 51.0778$^\circ$ & 0.3 & 0.6  & 8.5 & 7.7  & 6.0  & 4.8  & 166 & 0.06 & S & P206+52 & 4 & 0.00 & $\times$ \\
+510440.1 & $\pm$0.6$\arcsec$ & $\pm$0.7$\arcsec$ & $\pm$0.1 & $\pm$0.1 & $\pm$1.8 & $\pm$1.5 & $\pm$1.8 & $\pm$1.5 & $\pm$87 \\
ILTJ150957.26 & 227.4886$^\circ$ & 54.5732$^\circ$ & 0.4 & 1.9  & 15.8 & 11.4  & 14.6  & 9.7  & 99 & 0.14 & S & P227+53 & 2 & 0.00 & $\times$ \\
+543423.6 & $\pm$2.5$\arcsec$ & $\pm$1.6$\arcsec$ & $\pm$0.1 & $\pm$0.2 & $\pm$5.9 & $\pm$3.8 & $\pm$5.9 & $\pm$3.8 & $\pm$56 \\

 \hline  
 \end{tabular}
\end{sidewaystable*}

\section{Image quality}
\label{sec:image_quality}

The observations used in this data release were conducted between 2014
May 23 and 2015 October 15 and the varying observing conditions
significantly impact the image quality even after direction-dependent
calibration, which reduces the impact of ionospheric disturbances.
In this section, we assess the derived source sizes, astrometric precision, flux-density uncertainty, dynamic-range limitations, sensitivity, and
completeness, and briefly discuss some remaining calibration and imaging artefacts.

\subsection{Source extensions}
\label{sec:source_extensions}

Identifying unresolved sources using the PyBDSF-derived measurements
is complicated by several factors. For example, astrometric errors in
the mosaiced images cause an artificial broadening of sources, the
varying quality of calibration blurs the sources by differing amounts,
time averaging and bandwidth smearing can artificially extend sources, and the extent to which a source is deconvolved impacts its measured size. To accurately quantify all this would require realistic simulations in which compact sources are injected into the $uv$-data taking into account DDEs. Furthermore, as the precise criteria for distinguishing resolved sources varies between facets and observations, a prohibitively large number of these simulations would need to be performed. Our calibration and imaging pipelines are continuing to evolve and hence such a large undertaking is beyond the scope of this present study. An alternative approach would have been to inject point sources into our maps and use these to characterise the source finding algorithm; however, such a simulation would not account for distortions in source morphologies caused by calibration inaccuracies. Instead we attempted to assess whether or not sources are resolved by looking at the extensions of real sources that we assert are unresolved and we used these to define an average criterion with which additional unresolved sources can be identified across the entire mosaic.

To create a sample of unresolved sources the LoTSS-DR1 catalogue was first filtered to contain only isolated sources, which we define as being sources with no other LoTSS-DR1 source within $45\arcsec$. Any sources that were not in the deconvolution mask in every pointing in which they are detected were also excluded. From the remaining entries we then selected only sources that are classified as `S' by PyBDSF; this source code corresponds to sources that are the only objects within a PyBDSF island and are well fit with a single Gaussian. Finally, as described below, we imposed a cut on the major axis of the LoTSS-DR1 sources to limit the maximum extent of the low-frequency emission.

We emphasise that, owing to imperfect calibration, most truly unresolved sources in the LoTSS-DR1
catalogue do not have an integrated flux density to peak brightness
ratio of 1.0 or a fitted major axis size of 6$\arcsec$ (i.e. a size equal to the restoring beam). For example, the approximately 50 seemingly compact, bright (signal to noise in excess of 500) sources that meet the above criteria all have measured sizes in the FIRST catalogue of less than 5$\arcsec$ and we can therefore assert these are either unresolved or barely resolved. However, in the LoTSS-DR1 catalogue these sources have a median ratio of the integrated flux density to peak brightness equal to 1.12 with a median absolute deviation of 0.04. Furthermore, for seemingly compact LoTSS-DR1 sources that are detected with a lower signal to noise there is significantly more variation in the measured integrated flux density to peak brightness ratio. To characterise this, and separate extended from compact sources, we derived }an envelope with the functional form $\rm{\frac{S_{int}}{S_{peak}} = 1.25+A\big(\frac{S_{peak}}{RMS}\big)^{B}}$ , which encompasses 95\% of the LoTSS-DR1 sources that meet the above criteria (see Fig.\ \ref{fig:sourcesizes}). The factor of 1.25 was derived from the median plus three times the median absolute deviation of the integrated flux density to peak brightness ratio of the seemingly compact high signal-to-noise ($\geq$500) sources. We used this envelope to define a boundary between compact and extended sources.

The fitted envelope is dependent upon the cut used on the major axis
of the LoTSS-DR1 sources and we explored the impact of this by varying
that selection criterion from 10$\arcsec$ to 20$\arcsec$ (see
Fig.\ \ref{fig:sourcesizes}). We find that this has little impact on
the classification of sources with signal to noise of more than 100 as
either extended or compact; however, it has a much larger impact on
sources with lower signal-to-noise ratios. Whilst there is no definite
value to use for this cut, we chose a 15$\arcsec$ limit on the
LoTSS-DR1 major axis, which gives a best fit envelope of
$\rm{\frac{S_{int}}{S_{peak}} =
  1.25+3.1\big(\frac{S_{peak}}{RMS}\big)^{-0.53}}$. There are a total
of 280,000 LoTSS-DR1 sources within this envelope and we define these
as compact. As a cross check we note that 19,500 of these sources
correspond to entries in the FIRST catalogue and in that catalogue
88\% of them are less than 5$\arcsec$ in size, indicating that they are also compact at higher frequencies.

\begin{figure}   \centering
   \includegraphics[width=\linewidth]{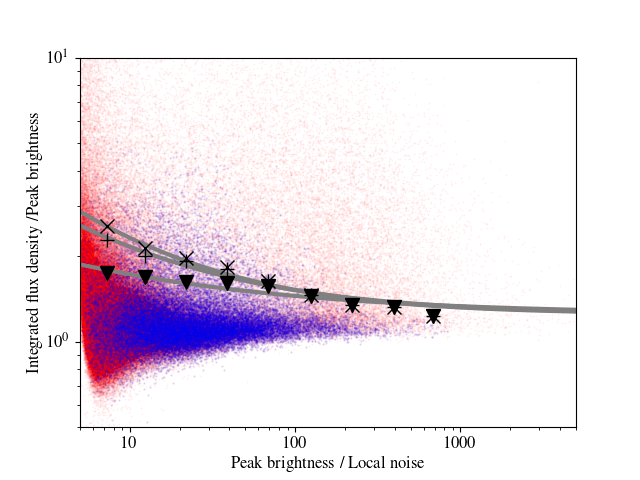}
   \caption{Ratio of the integrated flux density to peak
     brightness as a function of signal to noise for sources in the
     LoTSS-DR1 catalogue. All catalogued sources are shown in red and
     the sources we used to define an envelope that encompasses
     95\% of the compact sources are shown in blue (see
     Sec.\ \ref{sec:source_extensions}). The impact of varying the
     limit on the major axis size of LoTSS-DR1 sources is shown with
     the  triangles, crosses, and diagonal crosses corresponding to
     10$\arcsec$, 15$\arcsec$, and 20$\arcsec$ limits, respectively.
     Each of these is fitted with an envelope and the final selected
     envelope of $\rm{\frac{S_{int}}{S_{peak}} = 1.25+3.1\big(\frac{S_{peak}}{RMS}\big)^{-0.53}}$ was derived from the 15$\arcsec$ limit.}
   \label{fig:sourcesizes}
\end{figure}

\subsection{Astrometric precision}
\label{sec:astrometic}

The astrometry of our images is originally set by our phase calibration based on the TGSS-ADR1 catalogue. However, during direction-dependent calibration the astrometry can shift between regions because of the varying precision of the calibration models that are built up in different facets. For example, after direction-dependent calibration of a LOFAR data set \cite{Williams_2016} found $\sim1\arcsec$ offsets that varied systematically across their field, but they were able to correct these using the positions in the FIRST catalogue to provide a LOFAR HBA image with a standard deviation in the RA and Dec offsets from FIRST of just 0.4$\arcsec$. In our processing we also refined the astrometric accuracy after the self-calibration cycle is complete by correcting each facet independently using positions in the Pan-STARRS optical catalogue. Furthermore, during the mosaicing we do not include facets that have an uncertainty in the estimated astrometric correction of greater than 0.5$\arcsec$ to ensure high astrometric accuracy (see Sec.\ \ref{sec:data_reduction}).

To determine the resulting astrometric accuracy of our mosaic
catalogue we performed a simple nearest neighbour cross match in which
we took the closest Pan-STARRS, {\it WISE,} and FIRST counterpart that lies
within 5$\arcsec$ of each of the compact LoTSS-DR1 sources that were
identified using the procedure described in
Sec.\ \ref{sec:source_extensions}. We then created histograms of the
RA and Dec offsets and fit these with a Gaussian, where the location of
the peak and the standard deviation correspond to our systematic
position offset and the total uncertainty; these total errors are a
combination of errors in the LoTSS-DR1 positions from the source
finding software, the real astrometric errors in the LoTSS-DR1
positions, and the errors in the positions of objects in the cross-matched surveys
(which were selected owing to their high astrometric accuracy). The
astrometry of the Pan-STARRS catalogue was determined using a
combination of 2MASS and {\it GAIA} positions and the typical standard
deviation of the offsets from {\it GAIA} positions is less than 0.05$\arcsec$ (\citealt{Magnier_2016}). The {\it WISE} catalogue has a positional uncertainty of 0.2$\arcsec$  (\citealt{Cutri_2012}) in RA and Dec with respect to the 2MASS Point Source Catalog for sources detected at high significance, and the FIRST survey has astrometric uncertainties of 0.1$\arcsec$ with respect to the absolute radio reference frame (\citealt{White_1997}). 

\begin{figure}   \centering
   \includegraphics[width=\linewidth]{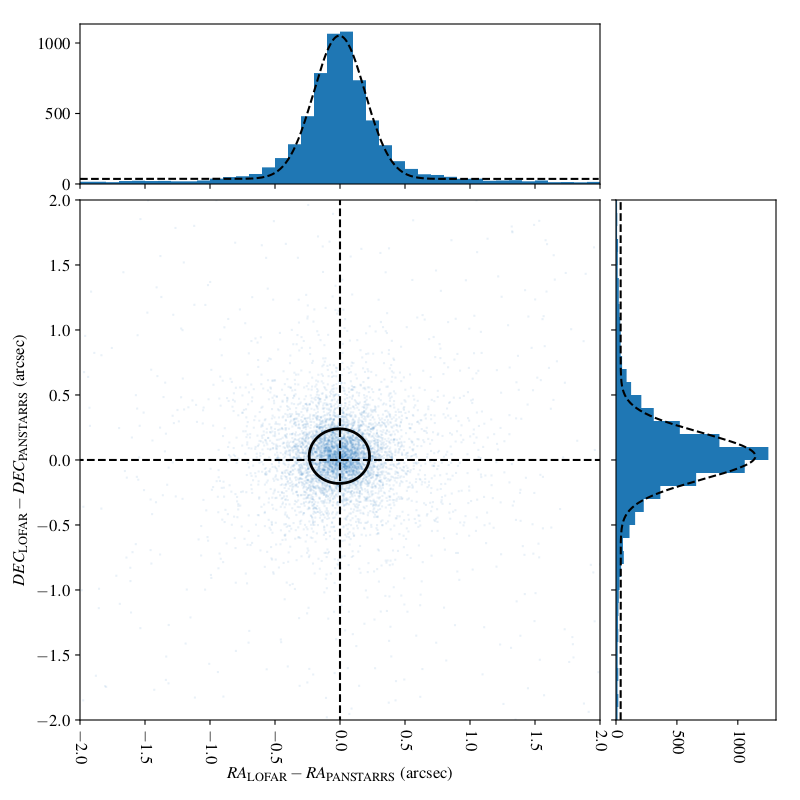}
   \caption{Residual RA and Dec offsets for LOFAR detected sources matched with their Pan-STARRS counterparts. The histograms show the number of sources at various RA and Dec offsets and the ellipse shows the peak location (less than 0.02$\arcsec$ from the centre in both RA and Dec) and the FWFM ($\sigma\approx0.2\arcsec$) of the Gaussian functions that are fitted to the histograms of the offsets. Similar plots showing the same LoTSS-DR1 sources cross-matched with {\it WISE} or FIRST sources show comparable systematic offsets and standard deviations of less than 0.27$\arcsec$ and 0.3$\arcsec$, respectively}
   \label{fig:first-astrometry}
\end{figure}

We cross-matched a total of 7100 sources from the LoTSS-DR1 catalogue to all three comparison sources and we found that, for these sources, there is a systematic positional offset from Pan-STARRS of less than 0.03$\arcsec$ and the standard deviation of the offsets is less than 0.2$\arcsec$ in both RA and Dec (see Fig.\ \ref{fig:first-astrometry}). Similarly, in comparison to {\it WISE}, we found the same sources have a systematic offset of less than 0.01$\arcsec$ and a standard deviation of less than 0.27$\arcsec$ in both RA and Dec. When comparing to FIRST, the systematic offsets are less than 0.02$\arcsec$ and the standard deviation is approximately 0.3$\arcsec$ in both RA and Dec. The direction of the derived systematic offsets varies when comparing the LoTSS positions with the three different surveys. We also examined the astrometric accuracy of our mosaic catalogue as a function of the LoTSS-DR1 peak brightness. We checked the accuracy of the catalogue to better estimate the real astrometric errors in the LoTSS-DR1 positions as bright ($\geq$20\,mJy), compact sources typically have errors in their derived positions of less than 0.05$\arcsec$. For the compact LoTSS-DR1 sources above 20\,mJy the fitted standard deviation to a Gaussian of the RA and Dec offsets from Pan-STARRS, and hence the approximate absolute astrometric accuracy of LoTSS-DR1, is less than 0.2$\arcsec$. The standard deviation gradually increases to $0.5\arcsec$ for the faintest LoTSS-DR1 sources ($\leq$0.6\,mJy) where the uncertainty in position from the source fitting can be as high as 1.0$\arcsec$. 

To assess the variation in the astrometric accuracy of various pointings prior to mosaicing the same analysis was performed on the catalogues derived from the LoTSS-DR1 images of the individual pointings. We only used similar sources to the previous analyses by first cross-matching the catalogues derived from the individual pointings with the LoTSS-DR1 compact source catalogue (see Sec.\  \ref{sec:source_extensions}). The resulting catalogue was then cross-matched with the Pan-STARRS catalogue. In addition we also imposed cuts on the catalogues from each LoTSS-DR1 pointing to include only sources within the 0.3 power point of the station beam, which is where the primary cut is made during the mosiacking. Furthermore, we only used sources classified by PyBDSF as `S' type sources in the pointing catalogues and those located in facets where the uncertainties in the Pan-STARRS dervied astrometric corrections of less than $0.5\arcsec$.  We found that the standard deviation of the Gaussian fitted to a histrogram of the RA and Dec astrometric offsets from Pan-STARRS varied from 0.31$\arcsec$ to 0.54$\arcsec$ with an average of 0.39$\arcsec$ and that the peak of the fitted Gaussian functions were displaced by between 0.05$\arcsec$ and 0.12$\arcsec$. These numbers give an indication of the varying astrometric accuracy across the HETDEX Spring Field region. We note that, as was found in the mosaiced images, these astrometric errors vary with the signal to noise of the detections and this explains why the  individual pointings have apparently larger astrometric errors than the mosaiced images. 

\subsection{Accuracy of the flux density scale}

Owing to inaccuracies in the existing LOFAR beam models, transferring
amplitude solutions derived from calibrators to the target field data
does not generally result in an accurate flux density scale for the
target field. For example, \cite{Hardcastle_2016} found the errors in the flux density scale to be up to 50\%. To correct this \cite{Hardcastle_2016} devised a bootstrapping approach to align the flux density scale of their LOFAR images with the flux density scales of other surveys whilst also providing more reliable in-band spectral index properties. We applied this technique early in the LoTSS-DR1 processing pipeline to ensure consistency with the VLSSr and WENSS flux density scales  (see Sec.\ \ref{sec:data_reduction}). To assess whether the flux density scale remains consistent throughout the processing we performed the same bootstrapping calculation with our final images. From our final images, the recalculated correction factors range from 0.8 to 1.3 with a mean of 1.0 and a standard deviation of 0.08. We did not apply these recalculated factors in this data release but they indicate that in some circumstances the flux density scale can drift during the processing; however, 60\% percent of the fields remain within 5\% of the original bootstrapped derived values. 

\begin{figure*}   \centering
   \includegraphics[width=0.45\linewidth]{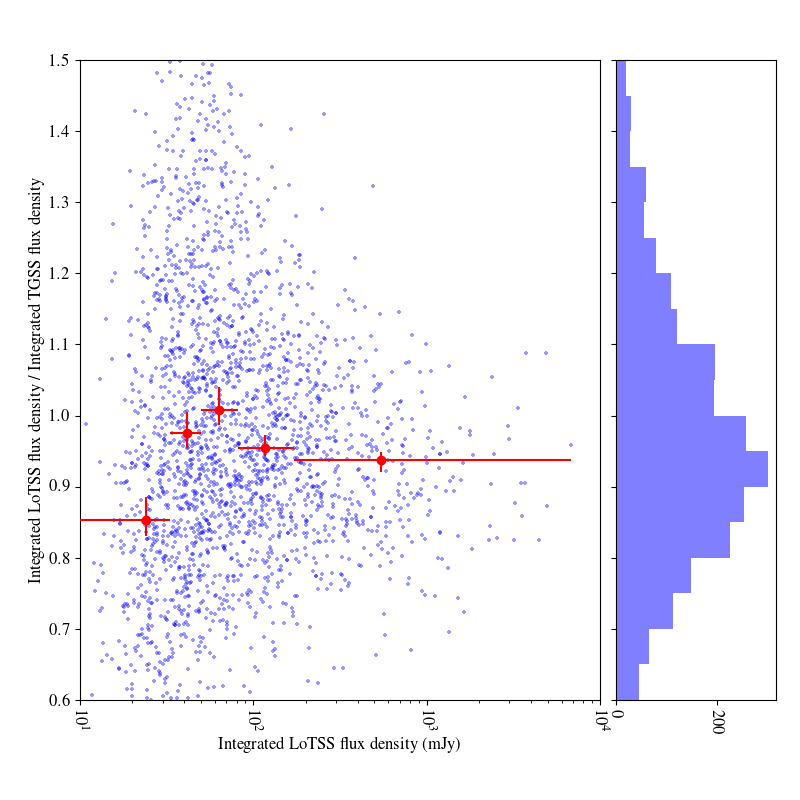}
   \includegraphics[width=0.45\linewidth]{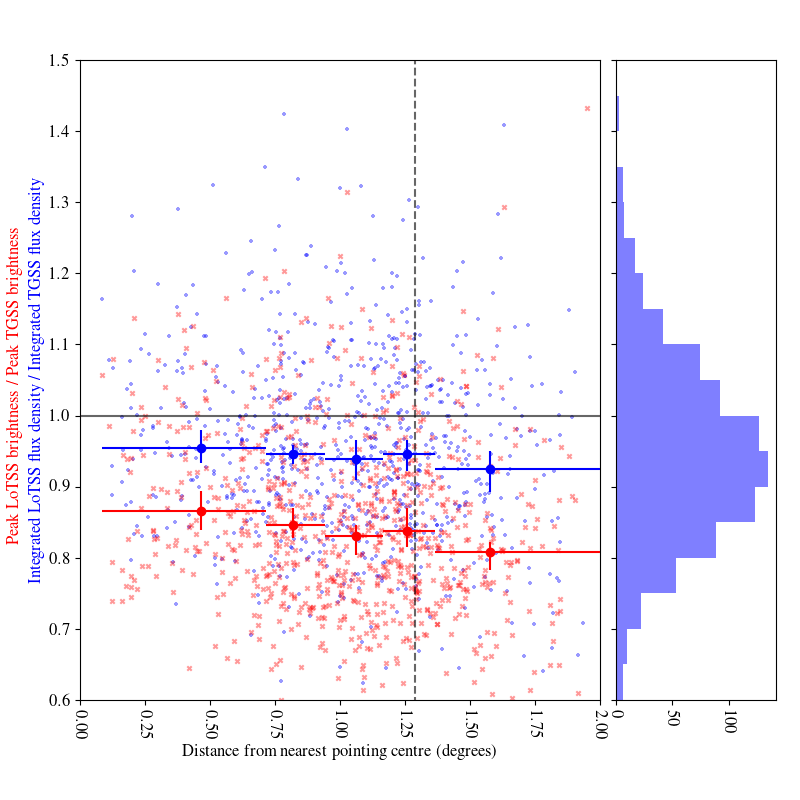}
  \caption{LoTSS-DR1 to TGSS-ADR1 integrated flux density ratio as a function of integrated flux density (left) and for sources with a integrated flux density higher than 100\,mJy as a function of distance from the nearest LoTSS pointing centre (right). Below 100\,mJy the completeness of TGSS-ADR1 drops below 90\% and, as a consequence, there is significant scatter in the integrated flux density ratio for sources below this limit. In the right panel we show that the 835 compact sources above this integrated flux density limit have a median integrated flux density ratio of 0.94 and a standard deviation of 0.14 (blue points) and a median peak brightness ratio of 0.83 and a standard deviation of 0.13 (red points). The thicker symbols show the median within bins indicated by horizontal error bars and the vertical error bars show the 95\% confidence interval of the derived median value estimated by the bootstrap method. The bins are chosen to contain equal numbers of sources, which is 500 and 170 for the left and right panels respectively. The vertical dashed line shows the median distance between LoTSS pointings and many of the measurements at greater distance are due to the edges in the LoTSS-DR1 mosaic.}
   \label{fig:tgssfluxratios}
\end{figure*}

For further verification of the flux density scale we compared the catalogued integrated flux density in the compact source LoTSS-DR1 catalogue to those in the TGSS-ADR1 catalogue. The TGSS-ADR1 measurements were not used during the bootstrapping to allow for this comparison. Furthermore, the TGSS-ADR1 flux density scale is not tied to the flux density scales of VLSSr or WENSS as the survey was instead calibrated directly against bright, well-modelled sources, on the \cite{Scaife_2012} flux density scale. For the 835 compact sources with LoTSS-DR1 integrated flux densities in excess of 100\,mJy the median ratio of the integrated LoTSS-DR1 flux densities to the integrated TGSS-ADR1 flux densities is 0.94 and the standard deviation of 0.14 (see the left panel of Fig.\ \ref{fig:tgssfluxratios}). However, at integrated flux densities below 100\,mJy, where the point-source completeness of the TGSS-ADR1 catalogue decreases to less than 90\% and detections are not always at very high significance, there is substantially more scatter in the ratio of TGSS-ADR1 to LoTSS-DR1 integrated flux densities with a standard deviation of 0.27.

Part of the scatter in the TGSS-ADR1 and LoTSS-DR1 integrated flux density ratios is from
variations in the quality of the images of various LoTSS-DR1
pointings. To examine the consistency of our measurements we compared
the integrated flux density of compact sources in catalogues derived
from each of the pointings used in LoTSS-DR1 with the TGSS-ADR1
catalogue. The median ratio of the LoTSS-DR1 integrated flux densities
to the TGSS-ADR1 integrated flux densities varies from 0.75 to 1.15
with an median of 1.0 and a standard deviation of 0.08. The discrepancy between this median integrated flux density ratio, which is derived from individual LoTSS-DR1 pointings, and corresponding value for the entire mosaic (0.94), appears to be a consequence of the mosaicing. Sources with apparently low LoTSS-DR1 integrated flux densities more often reside in pointings with apparently low noise levels that are more highly weighted during the mosaicing procedure. Furthermore,
we made use of the large overlap between pointings to examine
flux density scale variations and found that the standard deviation of
the median ratio of the integrated flux density between pointings is
0.2 and, whilst the maximum discrepancy in the integrated flux density
measurements is 55\%, 80\% of the ratios are within 20\% of unity. 

\begin{figure}   \centering
   \includegraphics[width=\linewidth]{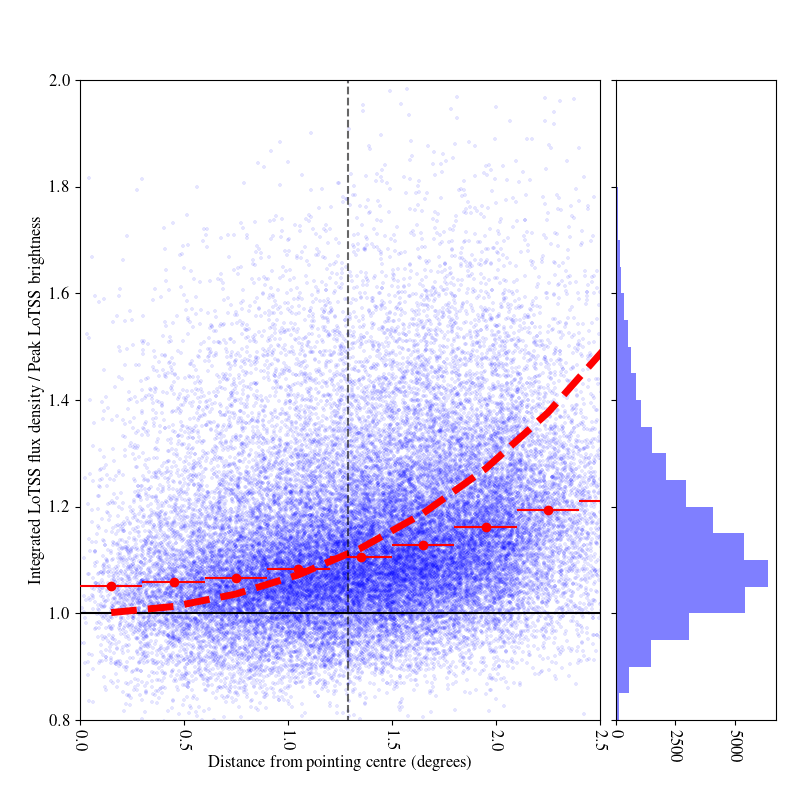}
   \caption{Integrated flux density to peak brightness ratios of compact LoTSS-DR1 sources as a function of separation from the pointing centre for catalogues derived from individual LoTSS-DR1 pointings. The thick red curve shows the approximate ratio expected from time and bandwidth smearing assuming that unresolved sources have a ratio of 1.0. The effects of time and bandwidth smearing are taken into account during deconvolution in \DDF/. The red points show the median ratios within bins of distance; the 95\% confidence intervals are $\sim$0.02 and were estimated by the bootstrap method. The horizontal
errors bars give the bin width and the vertical dashed line shows the median
distance between LoTSS pointings.} 
   \label{fig:fluxratiosradius}
\end{figure}

We also searched for trends between the source integrated flux density
measurements and the distance from the LoTSS pointing centres (see
Fig.\ \ref{fig:tgssfluxratios}). Using the 835 bright compact sources
in the mosaic catalogue that were cross-matched with TGSS-ADR1 we found
no strong dependence of the ratio of the LoTSS-DR1 integrated flux
density to the TGSS-ADR1 integrated flux density on the distance from
the closest LoTSS pointing;  the inner bin has a ratio of 0.95
and the outer has a ratio of 0.92. For the peak brightness the
radial dependence is slightly stronger with the inner bin at 0.86 and
the outer bin at 0.81. To assess the impact at further distances we
look at the peak brightness to integrated flux density ratio of compact sources
in the LoTSS-DR1 catalogues derived from individual pointings. Given that our data
are averaged to two channels per SB and 8\,s, it may be expected that
time-averaging and bandwidth-smearing effects are non-negligible in the LoTSS-DR1 mosaics; for example, we estimate using the formulas given by \cite{Bridle_1989} that at 6$\arcsec$ resolution the time-averaging and bandwidth smearing are as shown in Fig.\ \ref{fig:fluxratiosradius}. However, \DDF/ has a facet-dependent PSF which, for deconvolved sources, accounts for the impact of smearing. As a result the ratio of the peak brightness to integrated flux density in our LoTSS-DR1 images does not have as strong a dependence on distance from the nearest pointing centre as found in other studies that used imagers that do not correct for this. We note that there is still a small radial dependence. This may be because facets further from the pointing centre are generally  larger and, as a consequence, the ionospheric calibration in those regions is not as precise.
Overall, whilst there are variations in the accuracy of the flux density scale across the mosaic, we place a conservative uncertainty of 20\% on the LoTSS-DR1 integrated flux density measurements.

\subsection{Dynamic range}
\label{sec:dynamicrange}

The dynamic range in our images is limited and bright sources have an
impact on the image noise properties in a non-negligible fraction of the area that has been mapped. Whilst there are many factors that impact the dynamic range, our testing of the data processing procedure has indicated that the amplitude normalisation scheme that we  used certainly plays a significant role. Other contributors include the layout and size of the facets and the quality of the models that are built up during the self-calibration procedure.

\begin{figure*}
   \includegraphics[width=0.50\linewidth]{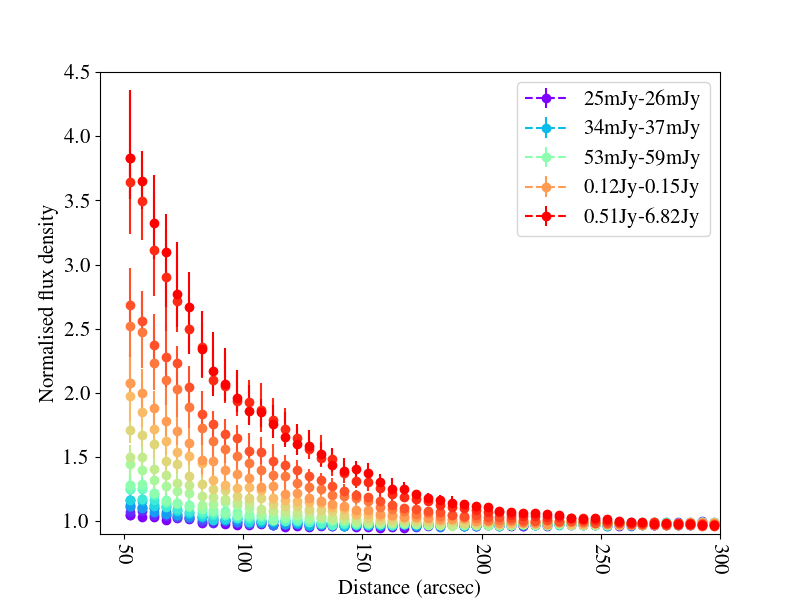}
   \includegraphics[width=0.50\linewidth]{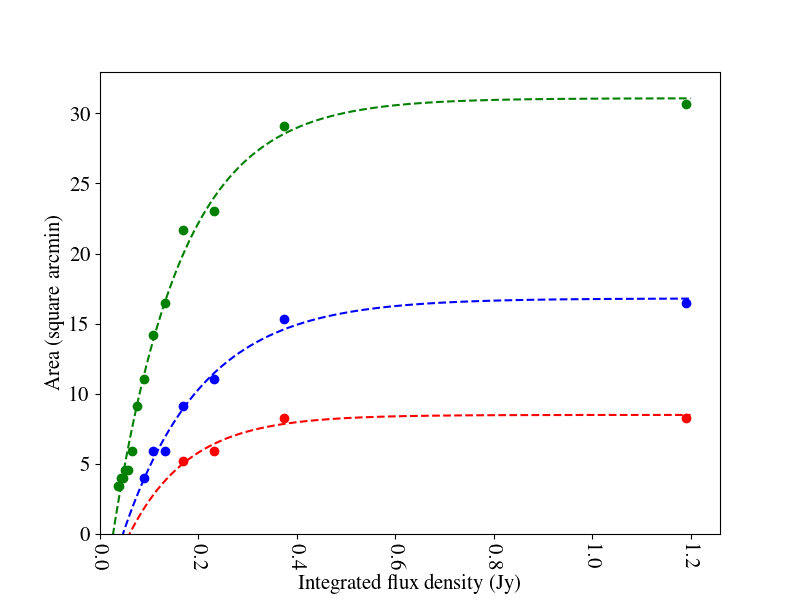}
   \caption{Left: Profiles of the average standard deviation as a function of distance from compact LoTSS-DR1 sources in various integrated flux density bins. The integrated flux density bins were chosen to contain equal numbers of sources (110) and they vary from 25.0\,mJy to 26.2\,mJy (blue) to 510\,mJy to 6820\,mJy (red). The errors are 95\% confidence intervals on the median values derived from the bootstrap method. Right: An estimate of the area in which the noise is 15\% (green), 50\% (blue), and 100\% (red) higher than the thermal noise due to dynamic-range limitations around a bright source; the points show the measured values and the dotted lines show the best-fitting curves.}
   \label{fig:dynamic-range}
\end{figure*}

To assess the dynamic-range limitations we examined pixels on mosaics of the final \DDF/ residual images in 5$\arcsec$ wide annuli around compact LoTSS-DR1 sources that were identified in Sec.\  \ref{sec:source_extensions}. A profile of the pixel standard deviation within every annulus was determined for each of these sources out to a radius of 500$\arcsec$. Each profile was fit with a Gaussian function plus a constant, which we assume is the level of the noise in the surrounding region and we used this to normalise the measurements. Within each distance bin, we averaged together all normalised noise measurements of sources within a given integrated flux density ranger and the mean and standard deviation was determined to create an average noise profile as a function of distance. These average noise profiles for various integrated flux density ranges are shown in Fig.\ \ref{fig:dynamic-range}.

The area in square degrees of sky that surrounds bright sources and has a noise level 
more than 15\% higher than the noise in the wider region depends
on the source integrated flux density according to approximately
$0.1(e^{-0.007S}-1)-0.002$, where $S$ is the integrated flux density
in mJy. From this equation, and removing overlapping regions, we
calculated that the noise is limited by the dynamic range of our maps
(i.e. the noise is more than 15\% higher than the noise level in
regions uncontaminated by bright sources) for 32 square degrees of the
424 square degrees that were imaged, i.e. 8\% of the total area
of the survey. Similarly, we calculated the area with even more enhanced noise levels of 50\% and 100\% higher than the noise level in uncontaminated regions as 3\% and 2\%, respectively.

\subsection{Sensitivity}

\begin{figure}   \centering
   \includegraphics[width=\linewidth]{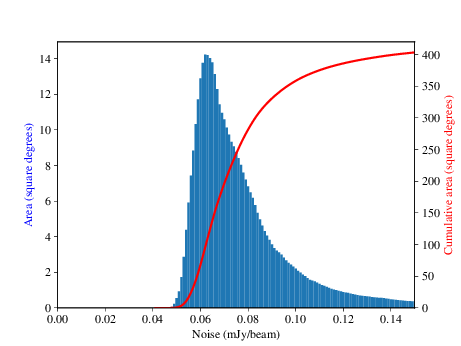}
   \caption{Estimated noise variations on the direction-dependent calibrated LoTSS-DR1 images. The red line shows the cumulative area of the mosaiced region that has an estimated noise less than a given value. The histogram shows the distribution of noise estimates within the mosaiced region.}
   \label{fig:mosaicdepth}
\end{figure}

The latitude of LOFAR is 52$^\circ$54$\arcmin$32$\arcsec$, putting the HETDEX Spring Field region, which has a declination ranging from 47$^\circ$ to 55$^\circ$, close to the optimal location where the projected area of the HBA dipoles and hence the sensitivity of the array is at its highest. The entire LoTSS-DR1 6$\arcsec$ resolution mosaic of the HETDEX Spring field region covers an area of 424 square degrees and the median noise level across the mosaic is  71\,$\mu$Jy beam$^{-1}$ ; 65\%, 90\%, and 95\% of the area has noise levels below 78\,$\mu$Jy beam$^{-1}$, 115\,$\mu$Jy beam$^{-1}$, and 147\,$\mu$Jy beam$^{-1}$, respectively (see Fig.\ \ref{fig:mosaicdepth}). These variations are due to varying observing conditions, telescope performance (e.g. missing stations or a higher level of interference), pointing strategy, and imperfections in the calibration and imaging procedure. The impact of the calibration and imaging procedure is particularly evident around bright sources in which the noise is limited by the dynamic range, as discussed in Sec.\ \ref{sec:dynamicrange}. The variations due to the observing conditions are also significant and the noise level on images of the individual pointings varies from 60\,$\mu$Jy beam$^{-1}$ to 160\,$\mu$Jy beam$^{-1}$. The sensitivity variations due to the mosaicing strategy in this region are much smaller. We find that the average mosaic noise as a function of distance from the closest pointing centre (just including regions covered by more than one pointing) only varies from 72$\mu$Jy beam$^{-1}$ to 78$\mu$Jy beam$^{-1}$ with a minimum at $\sim$1$^\circ$ from a pointing centre and a maximum at $\sim$1.6$^\circ$ from the nearest pointing centre. By comparison, the LoTSS-DR1 20$\arcsec$ resolution mosaic has higher noise levels due to the $uv$-cut applied in the imaging step. In this case the median noise level is 132\,$\mu$Jy beam$^{-1}$ , and 65\%, 90\%, and 95\% of the area has noise levels below 147\,$\mu$Jy beam$^{-1}$, 223\,$\mu$Jy beam$^{-1}$, and 302\,$\mu$Jy beam$^{-1}$, respectively.

The contribution of confusion noise to the total noise level that is measured on our 6$\arcsec$ resolution images is also small. To quantify this we followed the approach of \cite{Franzen_2016} and injected a broken power-law distribution of point sources convolved with a 6$\arcsec$ Gaussian into a blank image.  As in \cite{Franzen_2016} the  power law used for sources with an integrated flux density in excess of 6\,mJy was $\frac{dN}{dS} = 6998 S^{-1.54}$Jy$^{-1}$sr$^{-1}$ in agreement with Euclidean normalised differential counts at 154\,MHz derived by  \cite{Intema_2011},  \cite{Ghosh_2012}, and \cite{Williams_2013}. For fainter sources we fitted a power law of $\frac{dN}{dS} = 82 S^{-2.41}$Jy$^{-1}$sr$^{-1}$ to the deep 150\,MHz counts presented in \cite{Williams_2016} and, whilst these counts reach a depth of 700$\mu$Jy, for simplicity we assumed they hold to an integrated flux density limit of 10$\mu$Jy. Given that the counts are thought to decrease towards such low flux densities (e.g. \citealt{Wilman_2008}) this should result in a conservative estimate for the confusion noise. From the pixel values in the simulated image we derived the probability of deflection [P(D)],  which is highly skewed with an interquartile range of 18$\mu$Jy/beam. Whilst this distribution is not Gaussian, to approximate the confusion noise this can be converted to a crude estimate of the sigma by dividing the interquartile range by a factor of 1.349, which gives a confusion noise estimate at 6$\arcsec$ of 14$\mu$Jy/beam, which is significantly lower than the rms levels obtained. Our lower resolution images, however, are much more severely impacted by confusion noise and when repeating the analysis at 20$\arcsec$ our confusion noise estimate is 85$\mu$Jy/beam. We note that the very faint sources do not have a large impact on the sigma for the P(D) distributions; for example assuming the counts instead extend to 1$\mu$Jy assumes 5.1 million sources rather than 200,000 sources per square degree but increases the  20$\arcsec$ resolution confusion noise estimate by only 5\% to 89$\mu$Jy/beam. The power-law indices assumed in the calculations, however, play a more significant role; for example, again following \cite{Franzen_2016}, if for the sources between 10$\mu$Jy and 6\,mJy we assume $\frac{dN}{dS} = 6998 S^{-1.54}$, $1841 S^{-1.8}$, $661.8 S^{-2.0}$ or $237.9 S^{-2.2}$Jy$^{-1}$sr$^{-1}$ we estimate 20$\arcsec$ resolution P(D) sigma values of 1$\mu$Jy/beam, 10$\mu$Jy/beam, 24$\mu$Jy/beam, and 47$\mu$Jy/beam.

Several of the early LoTSS observations were conducted in a manner in which two neighbouring pointings were observed simultaneously, including 10 observations (thus 20 pointings) in this data release. In these circumstances a minor impact on the sensitivity in the overlapping regions of the simultaneously observed pointings is correlated noise. In an attempt to quantify the impact we examined pixel values in the overlapping regions of pointings by reprojecting the images to a common frame and ignoring regions containing sources (defined as those with values more than 1$\sigma$). The Pearson correlation coefficient calculated from these noise pixels was generally found to be 0.03-0.13 for pointings observed simultaneously but typically less than 0.03 for pointings observed at separate times. We also compared noise levels in mosaiced regions that contained data from two simultaneously observed pointings with regions where all contributing pointings were observed at different times. We found that regions where simultaneously observed pointings contribute to the mosaics have a median noise level that is $\sim$2\% higher than other regions.  The LoTSS observations of neighbouring pointings have not been conducted simultaneously since these very early observations.

\subsection{Completeness}
\label{sec:completeness}

To thoroughly estimate the completeness of the survey, sources of varying flux densities and positions should be injected into simulated data sets that include realistic DDEs. However, in the absence of such simulations, we instead inject sources into the direction-independent calibrated data sets taking into account the direction-dependent corrections that are applied in that specific direction to correct for the ionospheric and beam errors. After these data sets are processed with the pipeline and the injected sources are catalogued and their properties are compared to the parameters of the sources that were injected. We note that this procedure assumes that the direction-dependent corrections, with which the fake sourcs are injected, accurately describe the real DDEs. Given the computational cost of our calibration and imaging and that our pipelines will be improved for future data releases, we  only performed this analysis for 10 SBs of data from one pointing following the procedure outlined below:

\begin{enumerate}[label=\textbf{Step.\arabic*},ref=\textbf{Step.\arabic*},leftmargin=1.2cm]
\item Obtain the final direction dependent calibration solutions from a 240 SB run of the \LOTTSpipe/ pipeline
\item Create a simulated image of 300 delta functions drawn from a power-law distribution ($\frac{dN}{dS} \propto S^{-1.6}$)  and use \DDF/ to predict the visibilites for this model, corrupted by the same direction dependent distortions and add these to real direction independent calibrated data in the 10-SB data set
\item \label{complete:10sb} Execute the \LOTTSpipe/ pipeline on the 10-SB simulated data set
\end{enumerate}

\begin{figure}   \centering
   \includegraphics[width=\linewidth]{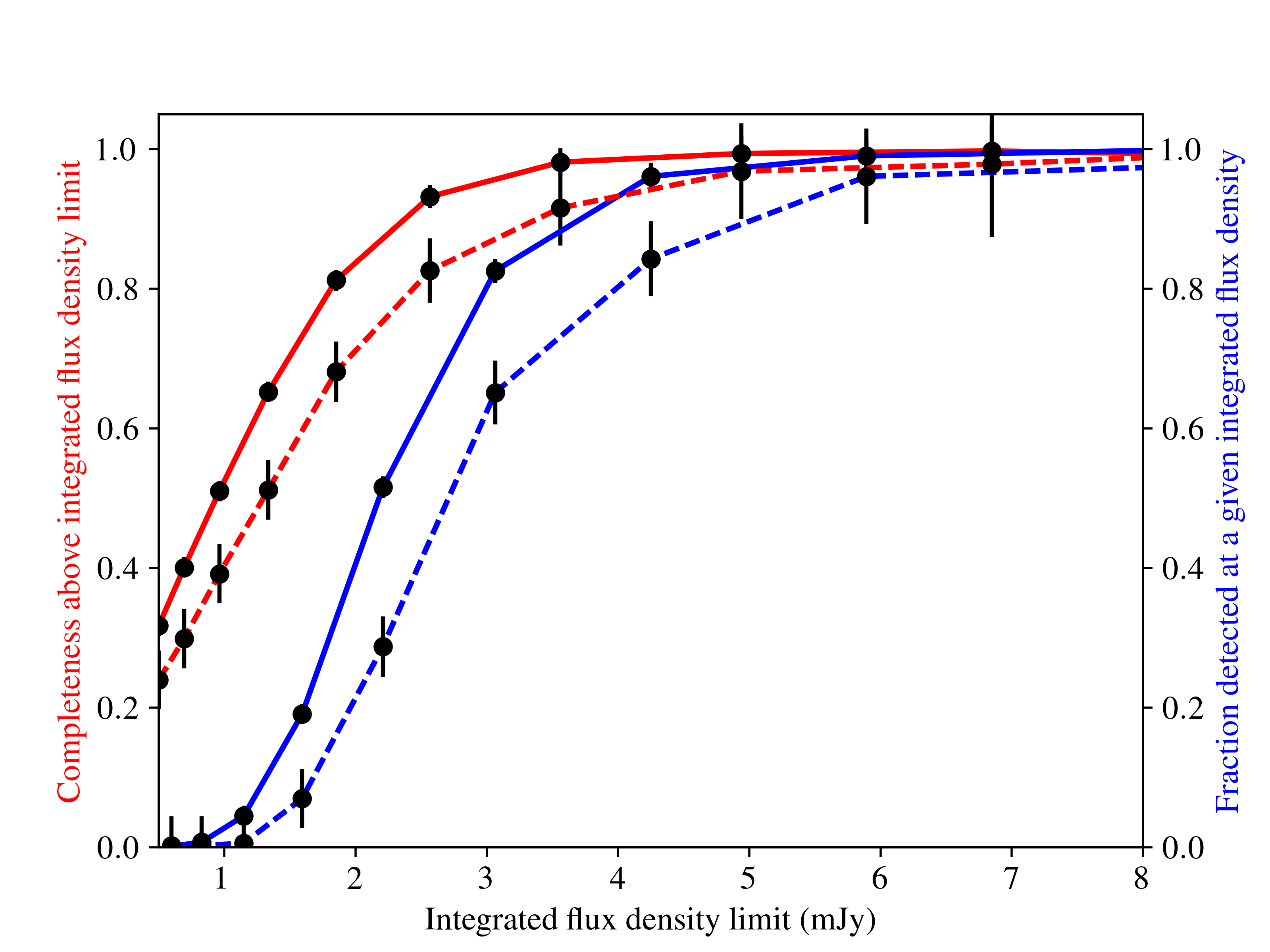}
   \caption{Estimated point-source completeness for a 10 SB (1/24th of the data) for a single LoTSS-DR1 pointing. The red line shows the completeness above a given integrated flux density and the blue line shows the fraction of sources detected at a specific integrated flux density value. The solid lines show the results of the simulation in which point sources are injected into PyBDSF residual images and the dashed lines show results from when delta functions corrupted by realistic direction-dependent errors are injected into the $uv$-data before it is run through \LOTTSpipe/. The error bars give the Poisson errors.}
   \label{fig:realcompleteness}
\end{figure}

For comparison, following the approach described in \cite{Heald_2015}, we also estimated the completeness by injecting 300 point-like sources with integrated flux densities drawn from a power-law distribution  ($\frac{dN}{dS} \propto S^{-1.6}$) into the final restored image of 1 of the 10 SB runs produced in \ref{complete:10sb}. To improve the statistics the realistic simulations were repeated 8 times giving a total of 2400 simulated point sources and the injection of sources into the final image was repeated 50 times giving a total of 15,000 sources. For both simulation types we ran PyBDSF on the simulated images and classifed the injected sources as detected if they are recovered within 7.5$\arcsec$ of the injected location and with a measured integrated flux density within 10 times the error on the integrated flux density uncertainty. The fraction of the simulated sources that were detected as a function of integrated flux density, and the derived completeness, for both methods are shown in Fig. \ref{fig:realcompleteness}. 

Whilst the injection of distorted point-like sources into the $uv$-data gives a much accurate understanding of the true completeness that we obtain from \LOTTSpipe/ it is computationally expensive to perform such simulations for the full bandwidth of each of the data sets in the survey with the full bandwidth of data. However, performing such simulations with 10 SBs of data from a single pointing suggests that the shape of completeness curves derived from realistic simulations is similar to that obtained from injecting sources into calibrated images (Fig. \ref{fig:realcompleteness}). Therefore,  to approximate the completeness of the entire LoTSS-DR1 we only used the less computationally expensive approach of injecting point sources into residual images. 

\begin{figure}   \centering
   \includegraphics[width=\linewidth]{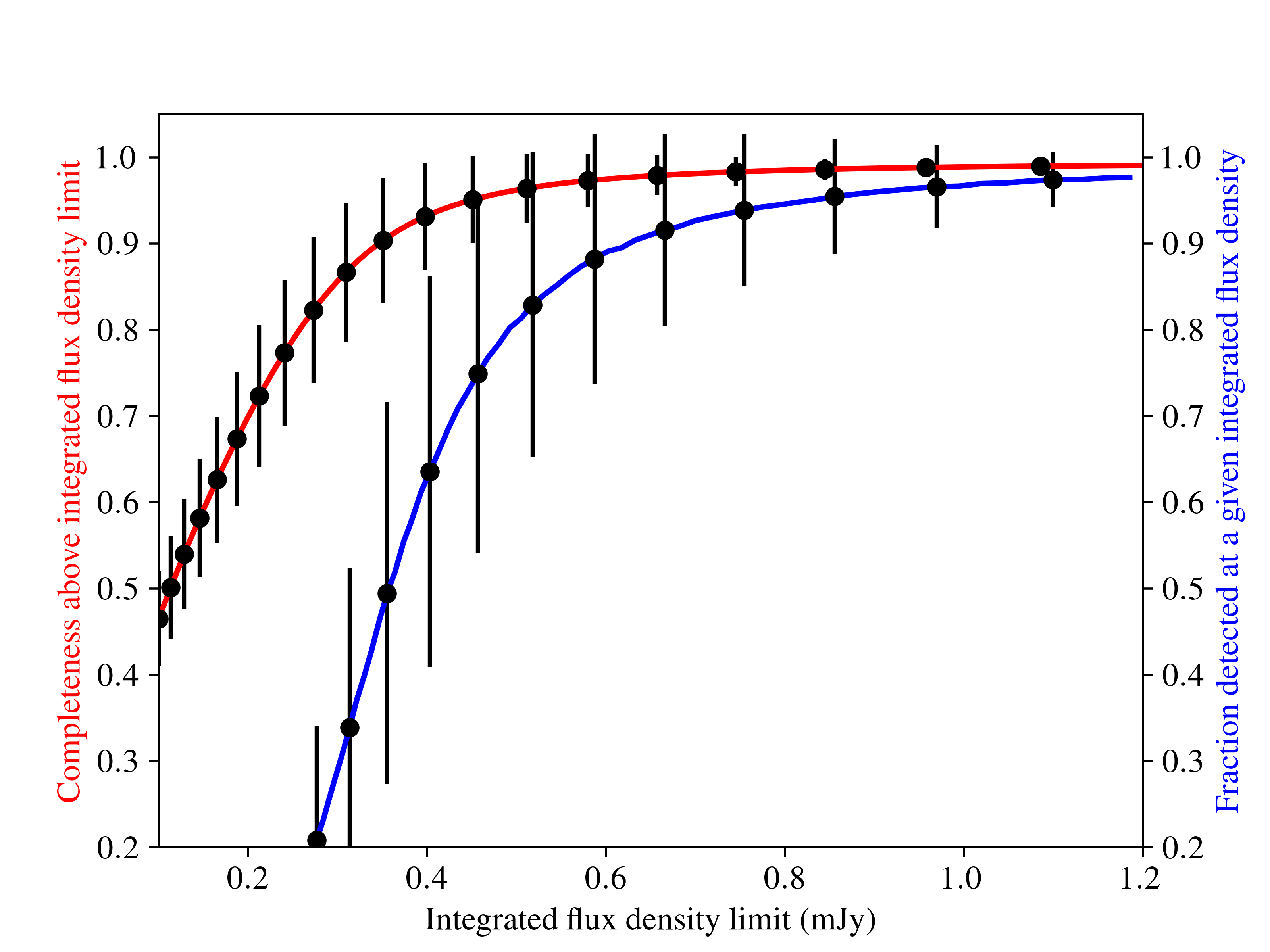}
   \caption{Estimated point-source completeness  of the LoTSS-DR1 catalogue. The red line shows the completeness above a given integrated flux density and the blue line shows the fraction of sources detected at a specific integrated flux density value. Because of the large number of sources injected during the simulation the Poisson errors are negligible but the errors bars reflect the standard deviation of the measurements as a function of position across the mosaic.}
   \label{fig:completeness}
\end{figure}

From each of the 58 mosaic images a residual image is generated using PyBDSF as a byproduct of the LoTSS-DR1 catalogue creation. Into each of these residual maps we inject 6,000 sources with integrated flux densities drawn from a power-law distribution ($\frac{dN}{dS} \propto S^{-1.6}$) and ranging from 0.1\,mJy to 10\,Jy. This procedure is repeated 50 times for each of the mosaiced images to ensure a statistically robust measurement.  The fraction of sources recovered above an integrated flux density limit, or the point-source completeness, varies with integrated flux density as shown in Fig.\ \ref{fig:completeness} and is 65\% at  0.18\,mJy, 90\% at 0.35\,mJy, and 95\% complete at 0.45\,mJy. However, we emphasise that, as shown in Fig. \ref{fig:realcompleteness}, the real integrated flux density level for the completeness levels is likely a factor of $\sim$1.3 higher (thus 90\% at 0.45\,mJy).

\subsection{Image artefacts}
\label{sec:artefacts}

In the LoTSS-DR1 mosaics there are several different types of artefacts.  The
low-level positive and negative haloes are particularly prominent around some sources; these haloes can be difficult to distinguish from real emission and make it
challenging to precisely characterise faint diffuse emission. These
artificial haloes are believed to be a product of having a minimum $uv$-distance on the
baselines used in the calibration; we suspect this expedient,
implemented to avoid modelling out extended emission, can cause the
amplitude solutions of the antennas with more short baselines to
become slightly discrepant from the more remote
antennas. For comparison with our images, we note that several diffuse objects within the region covered by this data release have been processed using a different direction dependent calibration algorithm (Facet Calibration; \citealt{vanWeeren_2016a}). This procedure does not use a large minimum $uv$-distance in the calibration and the images do not suffer from artificial haloes; see the maps presented in, for example \cite{Bruggen_2018}, \cite{Savini_2018}, and \cite{Wilber_2018}.  

In some fields there are also clear amplitude
calibration artefacts that are primarily a consequence of the
amplitude normalisation scheme that we used during the
direction-dependent calibration. Some fields that were  observed in
bad conditions also have clear phase errors that are dependent on
both our calibration solution interval (1min) and the size of the facets. Finally, whilst we attempted to ensure that our masks encompass extended sources, there are instances in which faint diffuse emission has still not been fully deconvolved. 

As described in Sec.\ \ref{sec:reducing_artefacts}, in future data releases we plan on improving upon each of these issues. However, for this data release, to aid with the identification of artefacts, we provided mosaics of the final residual maps to accompany our deconvolved continuum images along with the artefact flag resulting from the source (dis-)association and host galaxy identification work of the companion paper  \cite{Williams_2018}.

\section{Public data release}
\label{sec:cataloging} 

In this section, we summarise the products that form the first LoTSS public data release, which are accessible on \url{https://lofar-surveys.org}. These products consist of the mosaiced images that have been described in this paper in addition to the catalogue that we derived from the direct application of \textsc{PyBDSF} to the mosaiced 6$\arcsec$ resolution  images. In some cases, \textsc{PyBDSF} does not perfectly represent the radio source 
population: large extended radio sources may be split across several different catalogue entries 
in the \textsc{PyBDSF} catalogue, or alternatively two closely separated but physically distinct radio 
sources may be merged into a single catalogue entry by \textsc{PyBDSF}. Therefore, to enhance the scientific value of the released LoTSS-DR1 catalogues we attempted to associate or deblend the catalogued components of radio emission 
into actual radio sources where necessary, and also to identify the optical counterparts of all
sources. If an optical counterpart has been located we also estimated
its photometric redshift. For completeness, these projects that add value to publicly released LoTSS-DR1 catalogue are briefly summarised below, but
for a full description see Papers III and IV in this series \citep{Williams_2018,Duncan_2018c}.

\subsection{Mosaics and raw \textsc{PyBDSF} catalogue}

We  released both the 6$\arcsec$ and 20$\arcsec$ resolution 120-168MHz mosaiced images that were created following the direction dependent calibration procedure described in Sec.\ \ref{sec:data_reduction}. These mosaics cover 424 square degrees in the region of the HETDEX Spring Field (see Fig.\ \ref{fig:mosaic-noisemap}) and have the quality shown in Figs.\ \ref{fig:mosaic-example} and \ref{fig:example-complex-sources} and described in detail in Sec.\ \ref{sec:image_quality}. We released 6$\arcsec$
and 20$\arcsec$ mosaiced residual images to help assess the reliability of the morphology of extended structures;  these images show the quality of deconvolution and properties of the background noise. 

The raw \textsc{PyBDSF} catalogue that was released was created from the 6$\arcsec$ resolution mosaiced images; this catalogue is described in Sec.\ \ref{sec:mosaicing}. This catalogue contains 325,694 radio sources, has a source density of 770 sources per square degree and a point-source completeness of 90\% at an integrated flux density of 0.45\,mJy (see Sec.\ \ref{sec:completeness}). To aid the interpretation of the catalogue completeness we released the \textsc{PyBDSF} derived noise maps of the 6$\arcsec$ mosaics.

\subsection{Source (dis-)association and optical counterparts}

For most radio sources the expected host galaxy position is well defined by the properties of the radio source and it is therefore appropriate to use a statistical method to identify the counterparts in Pan-STARRS and {\it WISE}. For this we employ a likelihood ratio method (e.g. \citealt{Richter_1975},  \citealt{deRuiter_1977} 
and \citealt{Sutherland_1992}). However, for a number of complex sources, such methods are  either not possible or unreliable, so we employ a human visual classification scheme based on the Zooniverse\footnote{\url{www.zooniverse.org}} framework. Sources in the raw \textsc{PyBDSF} catalogue are first 
sorted based on their catalogued characteristics and selected either for visual (dis-)association and identification or for likelihood ratio cross-matching by means of a decision tree. The details of how these decisions are made and full details of the likelihood ratio and visual classification methods are given by \cite{Williams_2018}. Using this procedure, counterparts were identified for 71\% of the radio sources. These source characteristics and visual inspection procedure are also very useful in flagging probable artefacts in the \textsc{PyBDSF} catalogue. Again, details are given by \cite{Williams_2018}, but the final column of Table~1 provides a flag highlighting the \textsc{PyBDSF} sources identified as probable artefacts based on that work.

\subsection{Photometric redshift estimation}

Knowing the redshift of a source is a fundamental requirement for extracting key physical properties from continuum radio observations, such as luminosity or physical size, and for understanding the host galaxy (e.g. its stellar mass). Although future optical spectroscopy campaigns such as WEAVE-LOFAR\footnote{{\url{http://www.ing.iac.es/weave/weavelofar/}}} \citep{Smith_2016} will target more than $10^6$ 150 MHz-selected sources and provide
high-precision spectroscopic redshifts and accurate source
classifications for a large portion
of the LoTSS population, existing spectroscopic redshifts, largely
from SDSS, are available only for a very small subset of sources. Therefore, photometric redshifts (photo-$z$s) are a vital method for identifying the physical properties of radio sources and we produced photo-$z$ estimates for all plausible counterparts in the combined Pan-STARRs/All-{\it WISE} catalogue that was used for host-galaxy identification in the previous section. Full details of the photo-$z$ estimation, which combines template-based and machine-learning estimates, are presented in a companion release paper (\citealt{Duncan_2018c}).

\section{Future prospects}
\label{sec:futureprospects}

In future data releases we will not only present maps from a significantly larger fraction of the sky,  but there is also active development to improve many aspects of the LoTSS data processing; in the survey we observed almost 20\% of the
northern sky and in this work we only presented 10\% of these data or 2\% of
the northern sky. For example, to tackle the large LoTSS data rates we are working with the LOFAR e-infra group to implement our direction-independent calibration pipeline on the Forschungszentrum J\"{u}lich and Pozna\'{n} LTA sites. Furthermore, the observatory is beginning to utilise Dysco compression (\citealt{Offringa_2016}) to reduce the size of the archived data sets by a factor of approximately four. To improve the accuracy of the direction-independent calibration pipeline, amongst other things, the accuracy of the derived amplitude and clock solutions are being increased. In the direction-dependent calibration and imaging pipeline there is significant work to improve the fidelity of the images and to implement the pipeline on the SURFsara Grid. To further enhance the scientific potential of our data products there is also active work to exploit the polarisation (e.g. \citealt{vanEck_2018}), wide fractional bandwidth, the longest baselines provided by the international stations, and to use the data for spectral line studies (\citealt{Oonk_2017}, \citealt{Salas_2018} and \citealt{Emig_2018}) and for searches for transient sources. 

Discussing all of these future prospects in detail is beyond the scope of this article; however, in the following subsections we provide some details on several prospects, namely improving the direction-dependent calibration and exploiting the fractional bandwidth of LoTSS.

\subsection{Reducing image artefacts}
\label{sec:reducing_artefacts}

\begin{figure*}   \centering
   \includegraphics[width=0.45\linewidth]{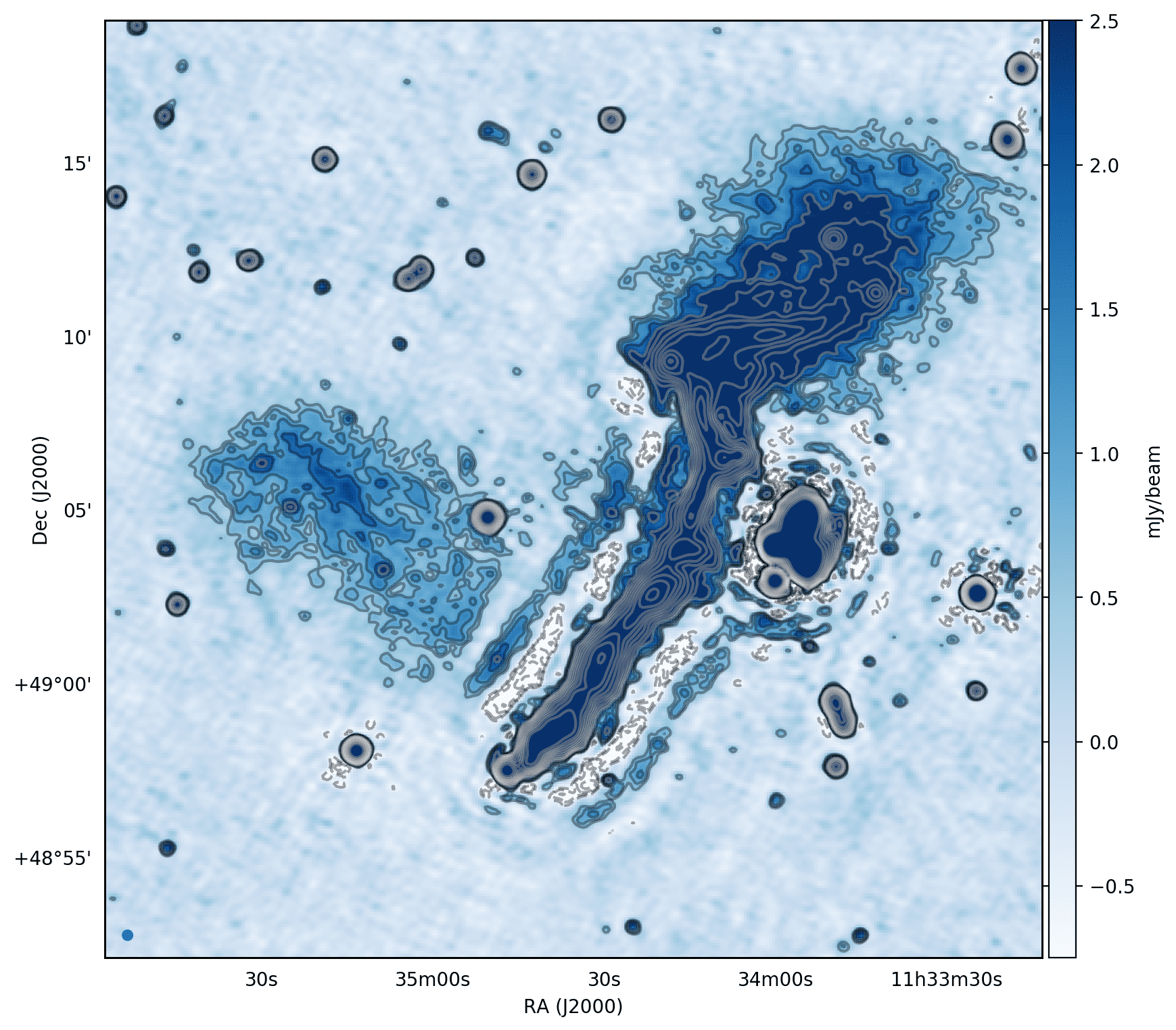}
   \includegraphics[width=0.45\linewidth]{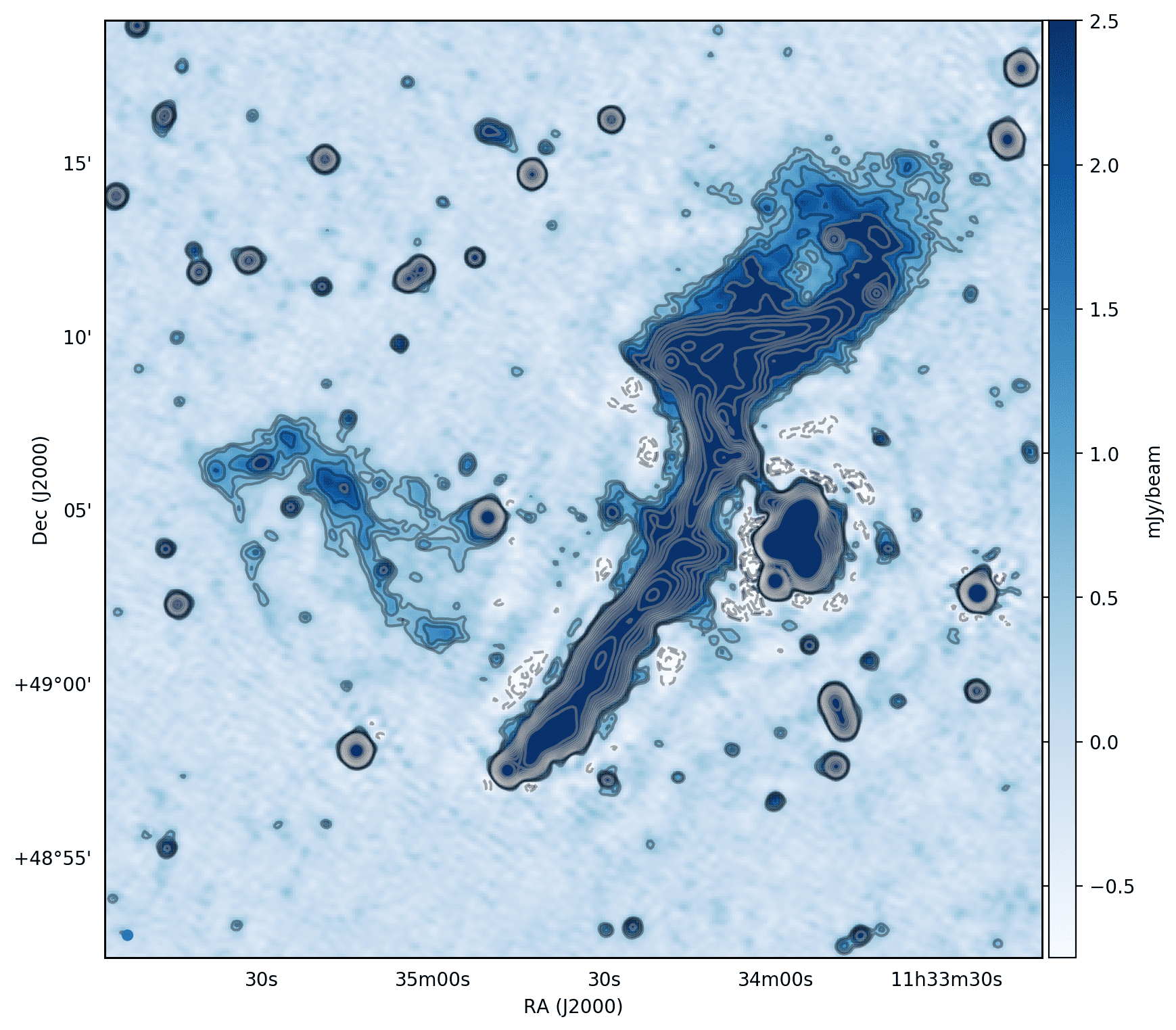}
   \caption{Left panel shows the 20$\arcsec$ resolution LoTSS-DR1 mosaiced image of the galaxy cluster Abell 1314 and right panel shows an image produced from the same data but after improvements to the pipeline described in Sec.\ \ref{sec:reducing_artefacts}. The colour scale of the images are the same and the contours show the $\pm\sqrt{1,2,4,...}\times 5 \sigma$ levels where $\sigma = 120\mu$Jy beam$^{-1}$. The LoTSS-DR1 image of this cluster suffers from artificial haloes around the extended structures and a low dynamic range. The improved LoTSS image has a higher fidelity and is in good agreement with the independently processed image presented in \cite{Wilber_2018b}.}
   \label{fig:DR2-v-DR1-example} 
\end{figure*}

The LoTSS-DR1 processing strategy has produced sensitive and good quality LOFAR images, however it failed for 8\% of the fields and, as described in Sec.\ \ref{sec:artefacts}, the final mosaics contain several different types of artefacts. Therefore, in an attempt to improve the images the development of the pipeline has been ongoing. The latest tests that use a refined recipe that still makes use of \kMS/ and \DDF/ for calibration and imaging, respectively, have shown that by removing the minimum $uv$-distance in the calibration and instead smoothing the amplitude solutions with a low-order polynomial function and fitting the phase solutions with a function proportional to $\nu^{-1}$ (which is, to first order, the phase behaviour introduced by free electrons in the ionosphere) the artificial haloes and holes can be effectively removed. Furthermore, these changes, together with other enhancements such as turning off the amplitude normalisation, improving the sky models used for calibration by increasing the depth of the deconvolution, and refining the direction-independent calibration by making use of accurate models derived from the direction-dependent imaging, have allowed us to decrease the failure rate of the pipeline, improve the dynamic range, and increase the number of sources detected. A demonstration of the improvements that are a result of these recent developments is shown in Fig.\ \ref{fig:DR2-v-DR1-example} and a refined version of the LoTSS processing pipeline will be fully described in a future publication.

\subsection{Exploiting the large fractional bandwidth of LoTSS}
\label{sec:fractional_bandwidth}

\begin{figure*}   \centering
   \includegraphics[width=0.425\linewidth]{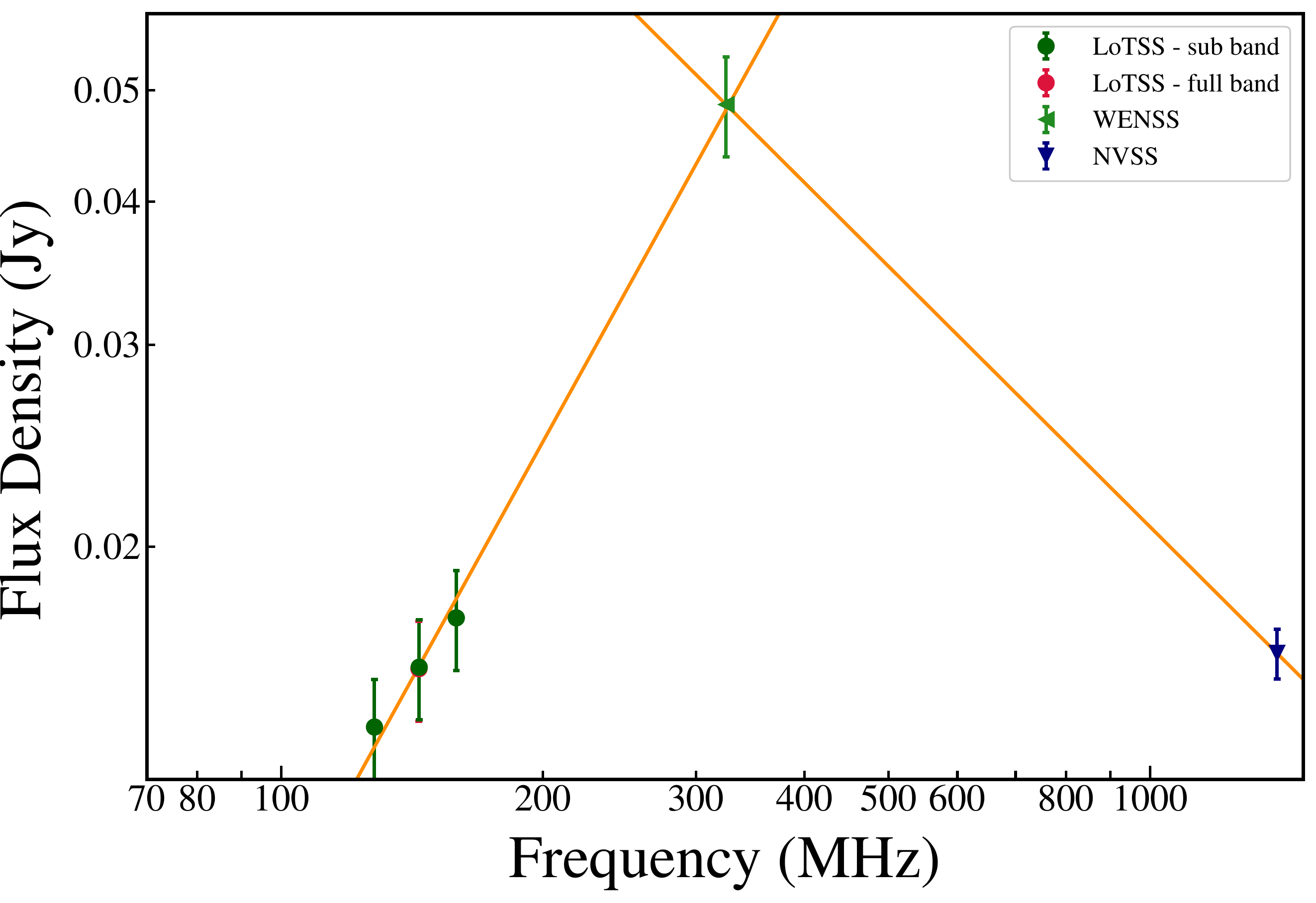}\hskip 1cm
   \includegraphics[width=0.419\linewidth]{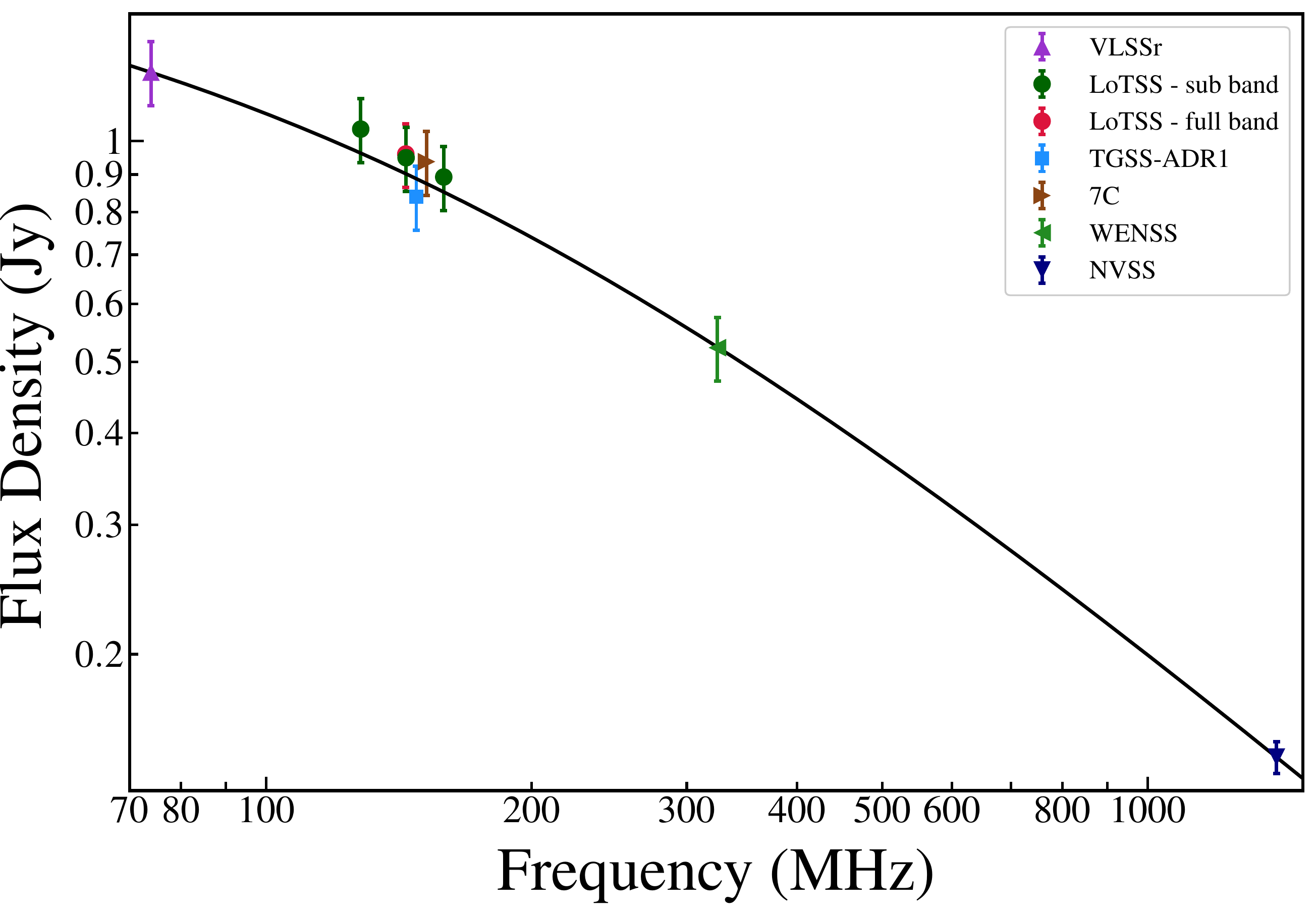}
  \caption{Example spectra showing the inband spectra of LoTSS. The left panel shows a low-luminosity peaked-spectrum source and the right panel shows a source that  deviates significantly from a power-law profile. The legend in each plot communicates the survey data used.
The best-fitting generic curved model fit (black) or power-law fit
(orange) is shown for each SED, derived from the model-fitting code described by \cite{Callingham_2015}.}
   \label{fig:examplespectra}
\end{figure*}

With a fractional bandwidth of approximately 33\%, LoTSS has the third largest 
fractional bandwidth of any very wide area radio continuum survey
produced to date. Only MSSS (\citealt{Heald_2015}) and the GLEAM (\citealt{Wayth_2015} and \citealt{Hurley-walker_2017}) 
survey have observed the sky with larger fractional bandwidths, but both have
significantly poorer angular resolutions and sensitivities (see Fig. \ref{fig:lotss_summary}). 
To demonstrate the scientific potential of the spectral information that can
be derived from LoTSS, for a test field we divided a direction-dependent calibrated LoTSS data set 
into three parts, each with a width of 16 MHz, and generated a
three-channel image with \DDF/. The integrated flux density measurements in each part of the bandwidth and the source association between the three images was done using
PyBDSF. An example of some observed spectra, with comparison to other surveys, is shown in Fig.\ \ref{fig:examplespectra}. In this demonstration field we were able to accurately derive (with 10\% uncertainty or less) in-band spectra for compact, isolated (no other source within 100$\arcsec$ of the LoTSS position) sources with integrated flux densities $\geq$ 10\,mJy, where the uncertainty estimate of the derived spectral indexes was obtained by comparing with spectral indexes measured from fitting to VLSSr and NVSS ($\geq$ 50\,mJy) or to TGSS-ADR1 and NVSS for the fainter sources ($\geq$ 10\,mJy).

Low-frequency spectral information is valuable for many science cases, such as for identifying low-luminosity peaked-spectrum sources (\citealt{Callingham_2017}) and investigating the energy distribution of electrons in the emitting region
of radio sources (e.g. \citealt{Bonavera_2011}). For example, the source shown in the left panel of Fig.\ \ref{fig:examplespectra} is a peaked-spectrum source with a radio luminosity $<10^{25}$\,W\,Hz$^{-1}$, which is two orders of magnitude fainter than the median radio luminosity of previous peaked-spectrum samples (e.g. \citealt{ODea_1998}). Probing this population of low-luminosity peaked-spectrum sources could potentially identify sources powered by a short-lived outburst of the central activity that might not able to escape from the host galaxy (\citealt{Czerny_2009}). Such sources could be the short-lived precursors needed to account for the overabundance of peaked-spectrum sources relative to the large-scale radio galaxies (\citealt{KunertBajraszewska_2010}). 

The right panel of Fig.\ \ref{fig:examplespectra} is the spectrum of a source that shows a significant deviation from a standard power law. If such a deviation is not taken into account, it leads to orders of magnitude uncertainty in the estimate of the energy stored by the lobes of the radio galaxy (\citealt{Duffy_2012,Harwood_2017}). 

Therefore, the spectral information that can be supplied by LoTSS will have diverse scientific impact, providing internal spectral index information to flux densities below the levels possible by cross-comparison with existing sky survey data. As a consequence of processing
constraints this spectral information is not included with this current release, but we plan to include it in future releases.

\section{Summary}
\label{sec:conclusions}

In this publication we have described the first full quality LoTSS
data release, which is available on-line \url{https://lofar-surveys.org}. We outlined how we managed the large LoTSS data
rate and we introduced the completely automated direction-dependent calibration and imaging pipeline that we used to produce 120-168\,MHz continuum images. The high-resolution (6$\arcsec$) images we present cover 424 square degrees in the region of the HETDEX Spring Field and contain 325,694 sources that are detected with a significance in excess of five times the noise. This source density is a factor of at least ten higher than any existing very wide area radio continuum survey. As described in companion papers (\citealt{Williams_2018} and \citealt{Duncan_2018c}) the LoTSS-DR1 catalogue has been enhanced by identifying the optical counterparts of radio sources and estimating their photometric redshifts. Finally, this data release is published together with $\sim$20 articles to highlight the scientific potential of LoTSS.

The LoTSS-DR1 images have a median sensitivity of 71$\mu$Jy beam$^{-1}$ with approximately 10\% of the mapped area being dynamic-range limited. For point sources, the survey is 90\% complete at a peak brightness of 0.45\,mJy beam$^{-1}$. 
We examined the fidelity of our images and found that the astrometric accuracy is approximately 0.2$\arcsec$ in both RA and Dec. The flux density scale is in overall agreement with other radio surveys and the uncertainty on the integrated flux density measurements is $\sim$20\%.

There are many opportunities to enrich the LoTSS data products through, for example
 polarimetric measurements or full exploitation of the longest
baselines in the international LOFAR array. We briefly demonstrated a few such possibilities including improvements to the calibration and imaging and the measurement of the in-band spectral index.

\section{Acknowledgements}
We thank the anonymous referee for his/her comments. This paper is based on data obtained with the International LOFAR
Telescope (ILT) under project codes LC2\_038 and LC3\_008. LOFAR
(\citealt{vanHaarlem_2013}) is the LOw Frequency ARray designed and
constructed by ASTRON. It has observing, data processing, and data
storage facilities in several countries, which are owned by various
parties (each with their own funding sources) and are
collectively operated by the ILT foundation under a joint scientific
policy. The ILT resources have benefited from the following recent
major funding sources: CNRS-INSU, Observatoire de Paris and Universit\'e
d'Orl\'eans, France; BMBF, MIWF-NRW, MPG, Germany; Department of Business, Enterprise and Innovation
(DBEI), Ireland; NWO, The Netherlands; The Science and Technology
Facilities Council (STFC), UK.

Part of this work was carried out on the Dutch national e-infrastructure with the support of the SURF Cooperative through grant e-infra 160022 \& 160152.  The LOFAR software and dedicated reduction packages on https://github.com/apmechev/GRID\_LRT were deployed on the e-infrastructure by the LOFAR e-infragroup, consisting of J. B. R. Oonk (ASTRON \& Leiden Observatory), A. P. Mechev (Leiden Observatory) and T. Shimwell (ASTRON) with support from N. Danezi (SURFsara) and C. Schrijvers (SURFsara).  

This work made use of the University of
Hertfordshire high-performance computing facility
(\url{http://uhhpc.herts.ac.uk}) and the LOFAR-UK computing
facility located at the University of Hertfordshire and supported by
STFC [ST/P000096/1]. The Data Center of the Nan\c{c}ay
Radioastronomy Station acknowledges the support of the Conseil
R\'egional of the R\'egion Centre Val de Loire in France. We thank Forschungszentrum J\"{u}lich for storage and computing.

This research made use of {\sc Astropy}, a community-developed core
Python package for astronomy \citep{AstropyCollaboration13} hosted at
\url{http://www.astropy.org/}, and of the {\sc astropy}-based {\it
  reproject} package
(\url{http://reproject.readthedocs.io/en/stable/}). We are grateful to
Thomas Robitaille for support in adapting the {\it reproject} package
to support the very large images used in this paper.


HR, DNH, KJD, and RJvW acknowledge support from the ERC Advanced Investigator programme NewClusters 321271.  
AB acknowledges support from the ERC-Stg DRANOEL, no 714245. 
AOC gratefully acknowledges support from the European Research Council under grant ERC-2012-StG-307215 LODESTONE. 
RM and MB gratefully acknowledge support from the European Research Council under the European Union's Seventh Framework Programme (FP/2007-2013) /ERC Advanced Grant RADIOLIFE-320745. 
%
RJvW acknowledges support from the ERC VIDI research programme with project number 639.042.729, which is financed by the Netherlands Organisation for Scientific Research (NWO). KLE acknowledges financial support from the Dutch Science Organization (NWO) through TOP grant 614.001.351.
%
%
MJH and WLW acknowledge support from the STFC ST/M001008/1. PNB and JS acknowledge support from the STFC ST/M001229/1. JHC and BM acknowledge support from the STFC ST/M001326/1 and ST/R00109X/1. 
CL, RK, RKC, and BW acknowledge support from STFC studentships.
LA acknowledges support from the STFC through a ScotDIST Intensive Data Science Scholarship.
LKM acknowledges the support of the Oxford Hinzte Centre for Astrophysical Surveys, which is funded through generous support from the Hintze Family Charitable Foundation. 
LKM is also partly funded by the John Fell Oxford University Press (OUP) Research Fund. 
GJW gratefully acknowledges support from the Leverhulme Trust. 
VHM thanks the University of Hertfordshire for a research studentship ST/N504105/1. 
%
%
%
AD acknowledges financial support from German Federal Ministry for Education and Research (BMBF, Verbundforschung, projects D-LOFAR III and IV, grants 05A15STA and 05A17STA). 
S.P.O acknowledges financial support from the Deutsche Forschungsgemeinschaft (DFG) under grant BR2026/23.
%
AG acknowledges full support from the Polish National Science Centre (NCN) through the grant 2012/04/A/ST9/00083. 
MJ acknowledges support from the Polish National Science Centre under grant no. 2013/09/B/ST9/00599. Grudziacka 5, 87-100 Toru\'n, Poland. 
MKB acknowledges support from the Polish National Science Centre under grant no. 2017/26/E/ST9/00216. 
%
IP acknowledges support from the INAF SKA/CTA PRIN project ''FORECaST". 
%
EB, MA, and OS are supported by the South African Research Chairs Initiative of the Department of Science and Technology and National Research Foundation.  
%
%
GGU acknowledges OCE Postocdoral fellowship from CSIRO. 
%
%
SM acknowledges funding through the Irish Research Council New Foundations scheme and the Irish Research Council Postgraduate Scholarship scheme. This publication has emanated from research supported in part by a research grant from Science Foundation Ireland (SFI) under the Grant Number 15/RI/3204.

\label{lastpage}
\end{document}